\documentclass[a4paper,11pt]{article}
\pdfoutput=1
\usepackage{graphicx}
\usepackage{fullpage}
\usepackage{jheppub} % for details on the use of the package, please
\usepackage{bbm}
\usepackage{amsfonts}
\usepackage{slashed}
\usepackage{subfigure}
\usepackage{array}
\usepackage{verbatim}
\usepackage{booktabs}
\usepackage{color, soul}
\usepackage{mathrsfs}
\usepackage{epsfig}
\usepackage{graphicx}% Include figure files
\usepackage{dcolumn}% Align table columns on decimal point
\usepackage{bm}% bold math
\usepackage{amsmath}
\usepackage{amssymb}
\usepackage{multirow}

\setstcolor{red}
\title{\sffamily Top-philic Scalar Dark Matter with a Vector-like Fermionic Top Partner}

\author[a]{Seungwon Baek,}
\author[a,b]{Pyungwon Ko,}
\author[a]{Peiwen Wu}

\affiliation[a]{School of Physics, KIAS, 85 Hoegiro, Seoul 02455, Republic of  Korea}
\affiliation[b]{Quantum Universe Center, KIAS, 85 Hoegiro, Seoul 02455, Republic of Korea}

\emailAdd{swbaek@kias.re.kr}
\emailAdd{pko@kias.re.kr}
\emailAdd{pwwu@kias.re.kr}

\abstract{We consider a simple extension of the Standard Model with a scalar top-philic Dark Matter (DM) $S$ coupling, apart from the Higgs portal, exclusively to the right-handed top quark $t_R$ and a colored vector-like top partner $T$ with a Yukawa coupling $y_{ST}$ which we call the topVL portal. 
When the Higgs portal is closed and $y_{ST}$ is perturbative $ (\lesssim 1)$, $TS\to (W^+b, gt)$, $SS\to t\bar{t}$ and 
$T\bar{T}\to (q\bar{q},gg)$ provide the dominant (co)annihilation contributions to obtain $\Omega_{\rm DM} h^2\simeq 0.12$  in light, medium and heavy DM mass range, respectively.
 However, large $y_{ST}\sim\mathcal{O}(10)$ can make $SS\to gg$ dominate via the loop-induced coupling $C_{SSgg}$ in the $m_S<m_t$ region. In this model it is the $C_{SSgg}$ coupling that generates DM-nucleon scattering in the direct detection, which can be large and simply determined by $\Omega_{\rm DM} h^2\simeq 0.12$ when $SS\to gg$ dominates the DM annihilation. The current LUX results can exclude the $SS\to gg$ dominating scenario and XENON-1T experiment may further test $y_{ST}\gtrsim 1$, and $0.5\lesssim y_{ST}\lesssim 1$ may be covered in the future LUX-ZP experiment. The current indirect detection results from Fermi gamma-ray observations can also exclude the $SS\to gg$ dominating scenario and are sensitive to the heavy DM mass region, of which the improved sensitivity by one order will push DM mass to be above 400, 600, 1000 GeV for $y_{ST}=0.3, 0.5, 1.0$, respectively.
$T\bar{T}$ pair produced at the hadron collider will decay $100\%$ into $t\bar{t}+\slashed{E}_T$ signal when kinematically open. The latest ATLAS 13 TeV 13.2 $\mathrm{fb^{-1}}$ data can excluded $m_T$ between 300
(650) and 1150 (1100) GeV for $m_S$ =40 (400) GeV and the exclusion region can reach up to $m_S\sim 500$ GeV.}

\begin{document}
\maketitle \indent
\newpage

\section{\label{introduction}Introduction}

The discovery of a new scalar particle at the Large Hadron Collider (LHC) whose properties are similar to those of the Higgs boson predicted in the Standard Model (SM) within the current experimental uncertainties was a huge success of particle physics community \cite{Chatrchyan2012,Aad2012}. However, the nature of Dark Matter  (DM) which occupies about $26\%$ of the current energy content of the Universe \cite{PlanckCollaboration2015-overview} is still a big puzzle. Since the SM cannot provide a suitable candidate for DM, many new physics models have been proposed to accommodate this new kind of matter.
The simplest extension of the SM would be a model with a singlet scalar DM $S$ which couples to the SM through the following Higgs portal (HP):
\begin{equation}
 {\mathcal L} \supset -\frac{1}{2} \mu_S^2 S^2 - \frac{1}{2} \lambda_{SH} S^2 H^\dagger H.
\label{eq:LSH}
\end{equation}
In the above Lagrangian a discrete $Z_2$ symmetry is assumed under which the DM is odd while the SM particles are even, which ensures 
stability of the DM. 
There are only two new parameters in this simple model, namely a DM mass parameter $\mu_S$ and a renormalizable 
quartic coupling $\lambda_{SH}$. The phenomenology of this simple Higgs portal model has been well studied (see for example \cite{Cline2013} and the references therein). The more-extended fermionic and vector Higgs portal models were studied in~\cite{Baek:2011aa,Baek:2012se}.

Apart from DM, top quark may also be a window to new physics beyond the SM. As the heaviest quark in the SM, it has the largest Yukawa coupling to the Higgs boson which implies it may play a special role in the electroweak symmetry breaking (EWSB). Top quark also provides the largest contribution to the running of
Higgs quartic coupling $\lambda_H$. A small change of top quark mass can significantly shift the energy scale where $\lambda_H$ becomes negative \cite{Degrassi:2012ry}, which makes the precise measurement of top quark properties very important for new physics studies at high energy scale.

Consequently, it is well motivated to connect the DM sector to top quark and a specially interesting scenario is the DM which couples only to the top quark sector. Some top-philic new particle sectors and/or DM models can be found in \cite{Kilic:2015vka,Beck:2015cga,Cox:2015afa,Berger:2011xk,Cheung:2010zf,Kumar2013-1303.0332,Kilic2015-1501.02202,Gomez2014-1404.1918,Bai2013-1308.0612,Agrawal2011-1109.3516} and the references therein. Different from previous studies, in this work we consider a top-philic scalar DM model by extending the above Higgs portal model with a vector-like fermionic particle $T$ (topVL) which is also odd under the unbroken discrete $Z_2$ symmetry. We require DM $S$ to couple to $T$ and the right-handed (RH) top quark $t_R$ via a Yukawa interaction with coupling $y_{ST}$. The Lagrangian including the Yukawa and new covariant kinetic terms reads:
\begin{equation}
 {\mathcal L} \supset \bar{T}(i \slashed D-m_T)T - (y_{ST} S \bar{T} t_R + h.c.) , 
\label{eq:LST}
\end{equation}
where $D_\mu$ is the SM covariant derivative. While the scenario of vector-like doublet coupling to the left-handed (LH) doublet $q_{3L}=(t_L,b_L)^T$ is completely analogous, in order to avoid any constraint from the bottom quark sector, we will focus on the $t_R$ case in this work. The gauge invariance requires the top partner $T$ to be also $\rm{SU(2)_L}$ singlet and have the same analyzed as $t_R$. Note that the $Z_2$-odd parity assigned to $T$ forbids it from mixing with the SM top quark, thus the current LHC constraints on heavy vector-like quarks do not apply here \cite{Aad:2015kqa,Khachatryan:2015axa}.

The above Yukawa interaction terms will generate DM annihilation $SS \to t\bar{t}$ through $t$-channel and the co-annihilations $TS,T\bar{T}$. Since the Higgs portal interaction shown in eq.(\ref{eq:LSH}) can also provide the $SS \to t\bar{t}$ process in the $s$-channel, there will be interference with the topVL portal which can be either constructive or destructive. As we will discuss later, when the Higgs portal interaction is closed by setting $\lambda_{SH}=0$, the topVL portal can be effective by itself to obtain the observed thermal relic density $\Omega_{\rm DM} h^2\simeq0.12$. However, interplay with the Higgs portal  can shift the topVL portal parameter space due to the interference in $SS \to t\bar{t}$ and the other annihilation channels provided by Higgs portal.

Another feature of this model is that $S$ can couple to the gluon via the 1-loop box diagram with $t$ and $T$ running inside \cite{Hisano2015}. This effective coupling $C_{SSgg}$ will provide the DM annihilation $SS\to gg$ and we found that large $y_{ST}\sim\mathcal{O}(10)$ can make 
$SS\to gg$ dominate in the $m_S<m_t$ region. Due to the absence of valence top quark in the nucleon, in this model it is the $C_{SSgg}$ coupling that generates DM-nucleon scattering in the direct detection (DD), which can be large and simply determined by $\Omega_{\rm DM} h^2\simeq 0.12$ when $SS\to gg$ dominates the DM annihilation. We found that the current LUX results can exclude the $SS\to gg$ dominating scenario and the expected sensitivity of XENON-1T may further test $y_{ST}\gtrsim 1$, and $0.5\lesssim y_{ST}\lesssim 1$ may be covered in the future LUX-ZP experiment.

The collider search for this model can be performed through the pair production $T\bar{T}$ which is dominated by the QCD processes. The top partner will decay $100\%$ into top quark and DM when kinematically open and produce $t\bar{t}+\slashed{E}_T$ signal which will receive constraints from the latest ATLAS 13 TeV 13.2 $\mathrm{fb^{-1}}$ data. We found that $m_T$ can be excluded between 300
(650) and 1150 (1100) GeV for $m_S$ =40 (400) GeV and the exclusion region can reach up to $m_S\sim 500$ GeV.

We note that a similar model was analyzed in Ref.~\cite{Giacchino:2015hvk} where DM couples only to light quarks $u_R$ or $d_R$.
Our model is phenomenologically distinguished from theirs in several aspects. For example, when the Higgs portal interaction is
turned off, the DM annihilation channel $SS \to q\bar{q} (q=u,d)$ is dominated by $d$-wave for $m_q \to 0$ in the light
quark portal model, while $s$-wave is allowed in our case.  Also in the absence of the Higgs portal, DM scattering off the nucleon occurs at the tree-level
in their case, while it occurs only via one-loop processes in our scenario.
The LHC signature in our model is also different from Ref.~\cite{Giacchino:2015hvk} which contains $SS \to 2j (t \bar{t})+\slashed{E}_T$, while ours is $t\bar{t}+\slashed{E}_T$. 
Different from Ref.~\cite{Giacchino:2015hvk} which neglected the Higgs portal interaction, we considered the interplay between the topVL and Higgs portal.
Other models where DM interacts with the SM leptons were considered in~\cite{Toma:2013bka,Giacchino:2013bta,Baek:2015fma,Baek:2015fea}.

This paper is organized as follows. In section \ref{section_RD} we study the various mechanisms in this top-philic scalar DM model to obtain the observed thermal relic density and the interplay among these mechanisms. In section \ref{section_DD} we investigate the $C_{SSgg}$ contribution to the DM direct detection through the loop process.  In section \ref{section_ID} we discuss the current constraints on this model from Fermi gamma-ray observations of dwarf galaxies and line spectrum. In section \ref{section_collider} we study the collider signal of this model based on the latest ATLAS 13 TeV 13.2 $\mathrm{fb^{-1}}$ data. We present the combined results in Section \ref{section_combine} and finally conclude in Section \ref{section_conclusion}.

\section{\label{section_RD}Thermal Relic Density}

The DM annihilation in this top-philic model can occur mainly via three different interactions: Higgs portal, topVL
portal and the effective $C_{SSgg}$ coupling. Since the Higgs portal mechanism has been well studied in other works, we
will first focus on the topVL portal by setting $\lambda_{SH}=0$. We also manually set $C_{SSgg}=0$ since, as we will
see later in section \ref{subsection_SSgg}, it is only effective with large $y_{ST}\gtrsim1$ and in the $m_S<m_t$ region
where the topVL portal is not sufficient. Then we turn on the $C_{SSgg}$ coupling to see its contribution to the DM
annihilation compared to $SS\to t\bar{t}$ and co-annihilations.
Finally we will bring the Higgs portal contribution back by setting $\lambda_{SH}=\lambda_0(m_S)\, r_\lambda$ where $\lambda_0(m_S)$ is the proper $\lambda_{SH}$ in the Higgs portal for $m_S$ to obtain the observed relic density, while $r_\lambda$ is some fractions such as $0.1, 0.2, 0.5, 1.0$ to control the Higgs portal strength. With these settings we are able to see the interplay between these two portals which can be either constructive or destructive in different parameter space.
%in which process there is no interplay between the former two portals due to the different annihilation products. (SS->TT, TT-> gg)
%Higgs invisible Br, lux, on Higgs portal

\subsection{\label{subsection_topVL}TopVL Portal}

DM in this model can annihilate via the topVL portal into $t\bar{t}$ final state. Both the DM and top partner can
co-exist in thermal equilibrium in the early Universe when $\Delta m=m_T-m_S \lesssim T_f$ with
$T_f$ the temperature at freeze out \cite{PhysRevD.43.3191}.  
This allows co-annihilations $TS,T\bar{T} \to \, \text{SM}$ which can become important when $SS\to t\bar{t}$ is
kinematically closed or not efficient.
We implemented this top-philic DM model with FeynRules
\cite{Alloul2014} and used micromegas \cite{Belanger2015,Belanger2014} to calculate the DM thermal relic density.

\begin{figure}[h]
\begin{center}
\includegraphics[width=12cm]{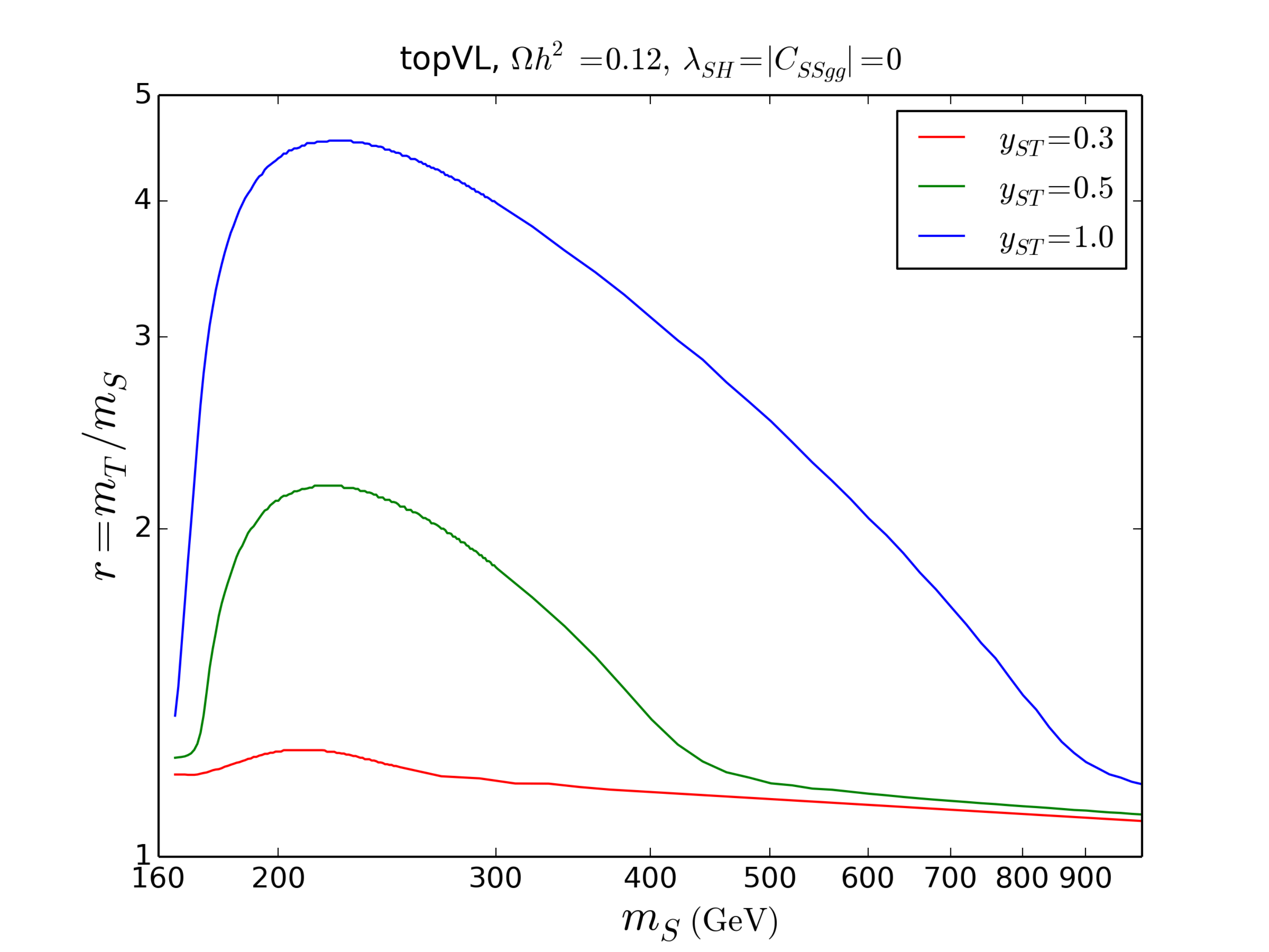}
\caption{Contour lines of $\Omega_{\rm DM} h^2=0.12$, in ($m_S$,$r$)-plane with $r=m_T/m_S$  for Yukawa
  coupling $y_{ST}$=0.3, 0.5, 1.0. Here we set $\lambda_{SH}, C_{SSgg}$=0. }
\label{fig:mdm_vs_r_topVL}
\end{center}
\end{figure}

In fig.\ref{fig:mdm_vs_r_topVL} we show contours with $\Omega_{\rm DM}  h^2=0.12$ which was measured by Planck
\cite{PlanckCollaboration2015}, in the plane of $m_S$ versus mass ratio $r=m_T/m_S$ for Yukawa coupling $y_{ST}$=0.3,
0.5, 1.0. Note that here in order to focus on the topVL portal exclusively, we have set $\lambda_{SH}, C_{SSgg}$=0. We
can see that for $m_S\lesssim m_t$ where $SS\to t\bar{t}$ is mostly below the threshold, $r=m_T/m_S$ must be close to 1
to annihilate efficiently. The co-annihilation processes become important as we can see from the fact that $r\approx
1$. However, larger $y_{ST}$ can alleviate this tension to some extent which can be seen for $m_S$=170 GeV where we have
$r$=1.2, 1.25, 2.0 for $y_{ST}$=0.3, 0.5, 1.0. When $SS\to t\bar{t}$ becomes kinematically open, the production of
on-shell $t\bar{t}$ can enhance the annihilation significantly. In order not to annihilate too fast the mass ratio $r$
in this case needs to deviate from 1 more than the $m_S<m_t$ case. This is especially apparent for larger $y_{ST}$
and for $m_S$=225 GeV we have $r$=1.25, 2.2, 4.5 for $y_{ST}$=0.3, 0.5, 1.0.
When DM mass becomes even heavier, the total annihilation cross section will receive overall suppression from the heavy propagator and/or smaller phase space, in which case the mass ratio $r$ also needs to be close to 1. In this regime the co-annihilation processes
become important again.  Again, larger $y_{ST}$ provides the topVL portal more room to cope with the suppression. For $y_{ST}$=0.3, 0,5, 1.0, it is not until $m_S$= 300, 450, 800 GeV that $r$ drops back to the value in the
$m_S<m_t$ range.

\subsection{\label{subsection_SSgg}TopVL and $C_{SSgg}$ coupling}

Now we study the effective $C_{SSgg}$ coupling between DM and gluon which has been calculated in
\cite{Hisano2015}. However, since the top quark mass is heavy, we should not use the approximated result in the limit
$m_t \ll m_S, m_T$. Instead, we used the full expression of $C_{SSgg}$  presented there. In the following we still turn off the Higgs portal
by setting $\lambda_{SH}=0$ and concentrate on how $C_{SSgg}$ contributes to the DM annihilation in the light DM mass
range.

\begin{figure}[h]
\begin{center}
\subfigure{\includegraphics[width=7.5cm]{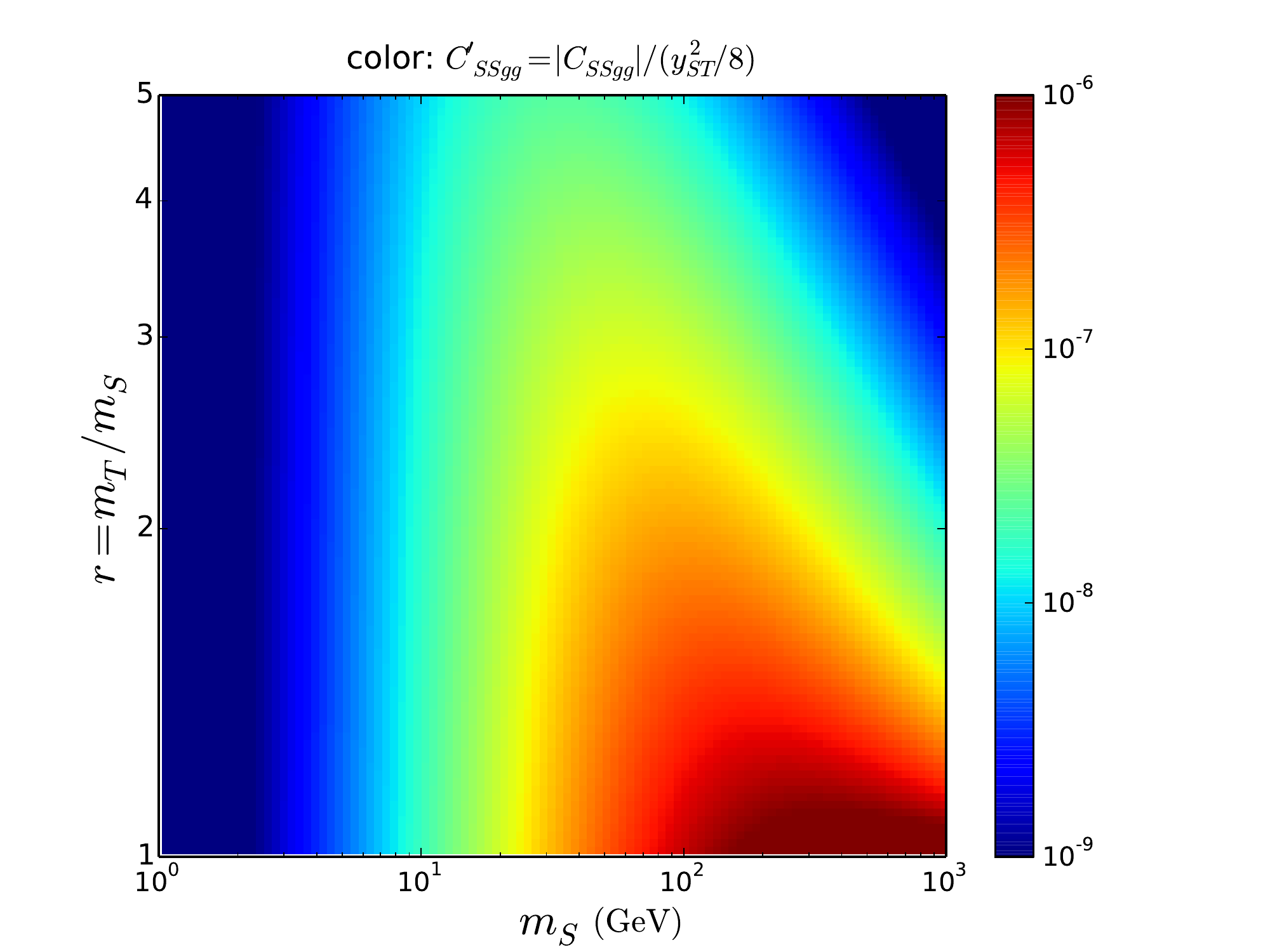}}
\subfigure{\includegraphics[width=7.5cm]{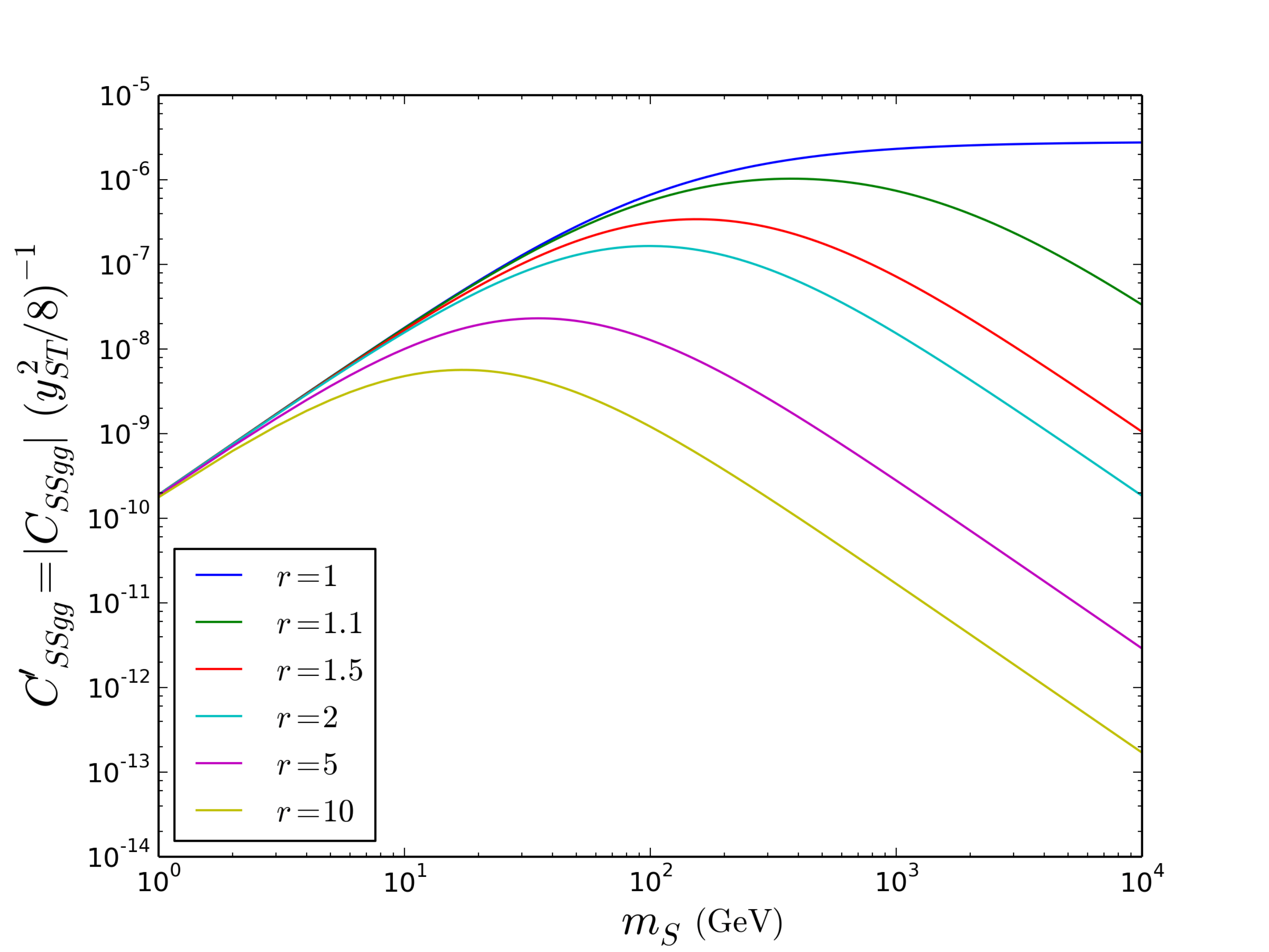}}
\caption{Value of $C'_{SSgg} \equiv|C_{SSgg}| (y_{ST}^2/8)^{-1}$ depending on $m_S$ and $r=m_T/m_S$. Left panel: on the same plane of $(m_S, r)$ as in fig.\ref{fig:mdm_vs_r_topVL}; Right panel: for fixed $r$=1.0, 1.1, 1.5, 2.0, 5, 10. Note that here we do not include the constraints from $\Omega_{DM} h^2=0.12$ and only show the general features of loop coupling $C_{SSgg}$.}
\label{fig:cssgg_prime}
\end{center}
\end{figure}

The value of $C_{SSgg}$ depends on $\{m_S, r, y_{ST}\}$ and in the limit
$m_t \ll m_S, m_T$ it has a simple expression $C_{SSgg} \approx -(y_{ST}^2/8)*[6m_S^2(r^2-1)^2]^{-1}$. The complete expression can be found in \cite{Hisano2015}. In the following, we extract the overall factor depending on $y_{ST}$ and define $C'_{SSgg} \equiv |C_{SSgg}| (y_{ST}^2/8)^{-1}$ and focus on the structure of $C'_{SSgg}$ with respect to $m_S$ and $r$. The left panel of fig.\ref{fig:cssgg_prime} shows how $C'_{SSgg}$ varies with $m_S, m_T$ on the same plane of $(m_S, r)$ as in fig.\ref{fig:mdm_vs_r_topVL}, while the right panel contains several fixed $r$=1.0, 1.1, 1.5, 2.0, 5, 10 for better understanding. Note that here we used the full expression of $C_{SSgg}$ since the DM mass region we consider include the case $m_S, m_T < m_t$. Moreover, fig.\ref{fig:cssgg_prime} does not include the constraints from $\Omega_{DM} h^2=0.12$ and only shows the general features of loop coupling $C_{SSgg}$.

The first thing one can notice is that $C'_{SSgg}$ is nearly independent of $r=m_T/m_S$ for very small $m_S$ and can be
very small $\sim10^{-10}$. For a fixed $r$, the value of $C'_{SSgg}$ will increase with increasing $m_S$ first and then
drop, except for the $r=1$ degenerate case where $C'_{SSgg}$ will approach a constant. Larger maximum
$C'_{SSgg}$ is obtained for smaller $r$ and the point where $C'_{SSgg}$ starts to drop occurs at larger $m_S$.
For a fixed DM mass $m_S$ the larger mass ratio $r$ will decrease $C'_{SSgg}$, especially for large $m_S$. These
features suggest that in fig.\ref{fig:mdm_vs_r_topVL} when $SS\to t\bar{t}$ is efficient, where DM mass is moderate and
$r$ is relatively large, $C_{SSgg}$ is generally suppressed and we checked that $SS\to
t\bar{t}$ in this region occupies almost $100\%$ of the DM annihilation (see fig.\ref{fig:fc} below). However, when $SS\to t\bar{t}$ is kinematically closed
or not efficient, $SS\to gg$ may play an important role, which can be more significant in the
$m_S<m_t$ region where $gg$ final state receives much smaller phase space suppression compared to co-annihilation. We should not forget that $y_{ST}^2$ is an overall factor in the full $C_{SSgg}$ which implies
that the curve with larger $y_{ST}$ in fig.\ref{fig:mdm_vs_r_topVL} can result in larger $SS\to gg$ contribution.

In fig.\ref{fig:fc} we show the contributions to the DM annihilation from different channels, for $y_{ST}=0.3, 0.5, 1.0,
10$. One can clearly see that the contribution from $SS\to t\bar{t}$ (green solid line) starts to dominate
the annihilation when kinematically open. With even heavier DM mass of several hundreds of GeV, it is the
co-annihilation channel $T\bar{T}\to q\bar{q},gg$ that dominates the contribution. However, larger
$y_{ST}$ can help $SS\to t\bar{t}$ dominate a wider DM mass range. As for the $m_S<m_t$ region, co-annihilations $TS\to
W^+b,gt$ have the largest contributions in most cases, while with larger $y_{ST}>1$ the
$SS\to gg$ (cyan solid line) can increase rapidly. In the extreme case with very large $y_{ST}=10$, $SS\to gg$ and
$SS\to t\bar{t}$ will dominate in most of the $m_S<m_t$ and $m_S>m_t$ region, respectively. This can be understood from the
fact that $SS\to gg$ depends on $y_{ST}^2$ while the co-annihilation $ST$ depends only on $y_{ST}$, which means $SS\to gg$ can
benefit more from large $y_{ST}>1$ than the co-annihilation $ST$. On the contrary, the contribution from gluon channel $SS\to gg$ is negligible with perturbative $y_{ST}\lesssim 1$, which means in this case the mass ratio $r=m_T/m_S$ is basically the same as those in fig.\ref{fig:mdm_vs_r_topVL} where we manually turned off gluon channel to show how the topVL portal itself generates $\Omega h^2=0.12$. In this case, for each point in fig.\ref{fig:fc}, one can estimate its loop coupling $C'_{SSgg}$ by comparing fig.\ref{fig:mdm_vs_r_topVL}  and fig.\ref{fig:cssgg_prime}, since a point with $\{m_S, r, y_{ST}\}$ read from fig.\ref{fig:mdm_vs_r_topVL} can be used to estimate its location in fig.\ref{fig:cssgg_prime} with $\{m_S, r\}$ and thus the corresponding $C'_{SSgg}$.

\begin{figure}[h]
\begin{center}
\subfigure{\includegraphics[width=7.5cm]{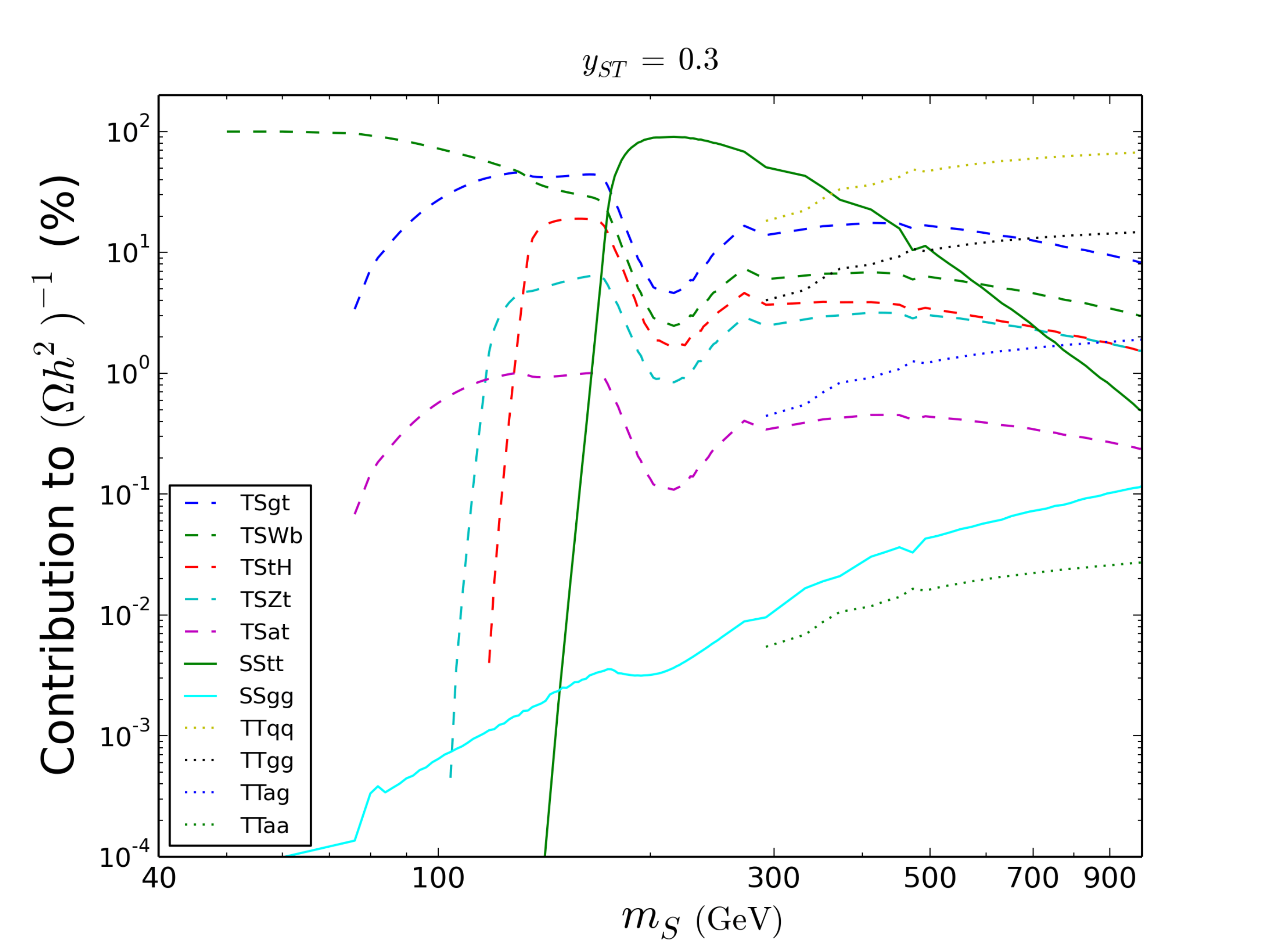}}
\subfigure{\includegraphics[width=7.5cm]{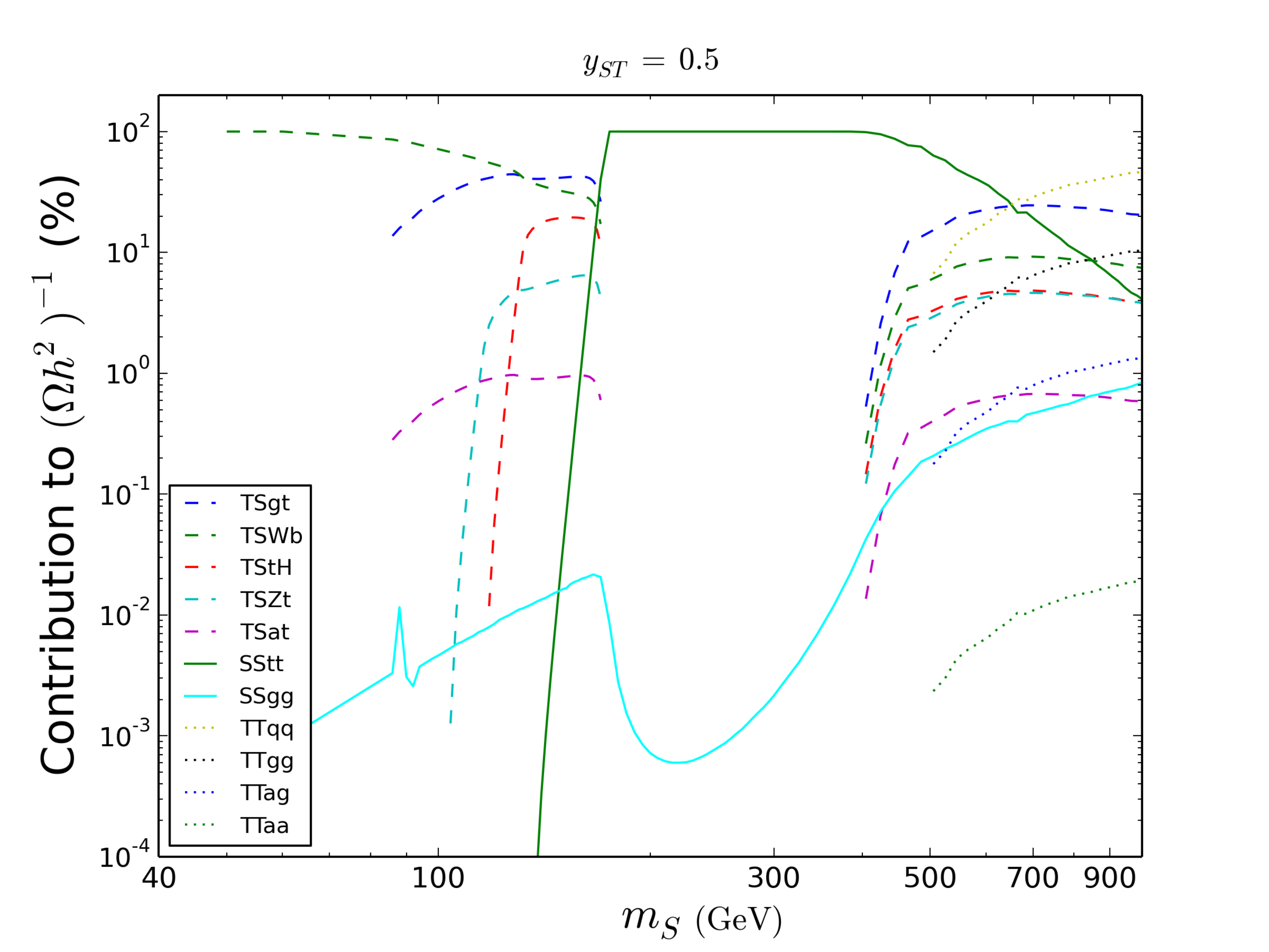}}
\subfigure{\includegraphics[width=7.5cm]{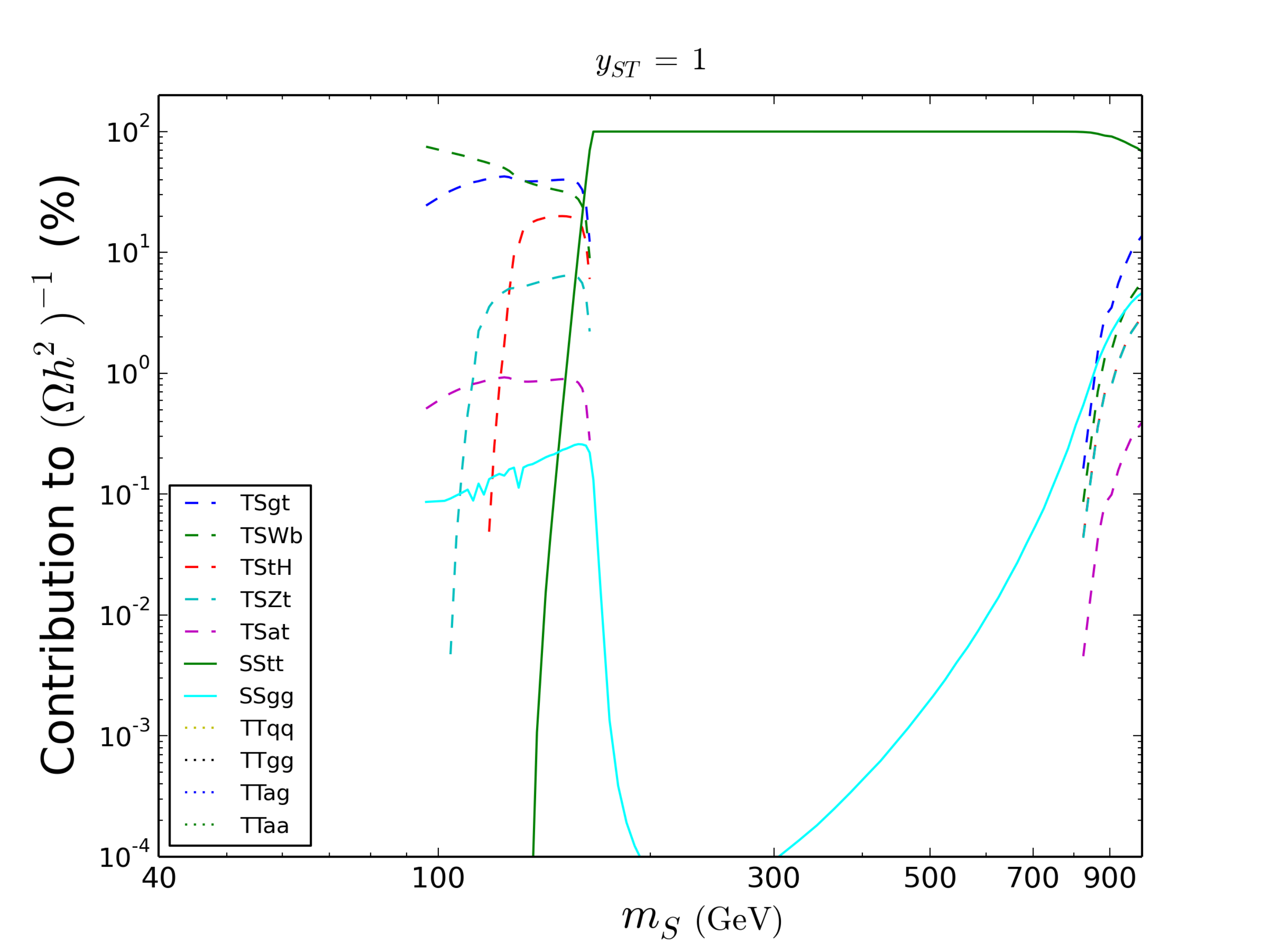}}
\subfigure{\includegraphics[width=7.5cm]{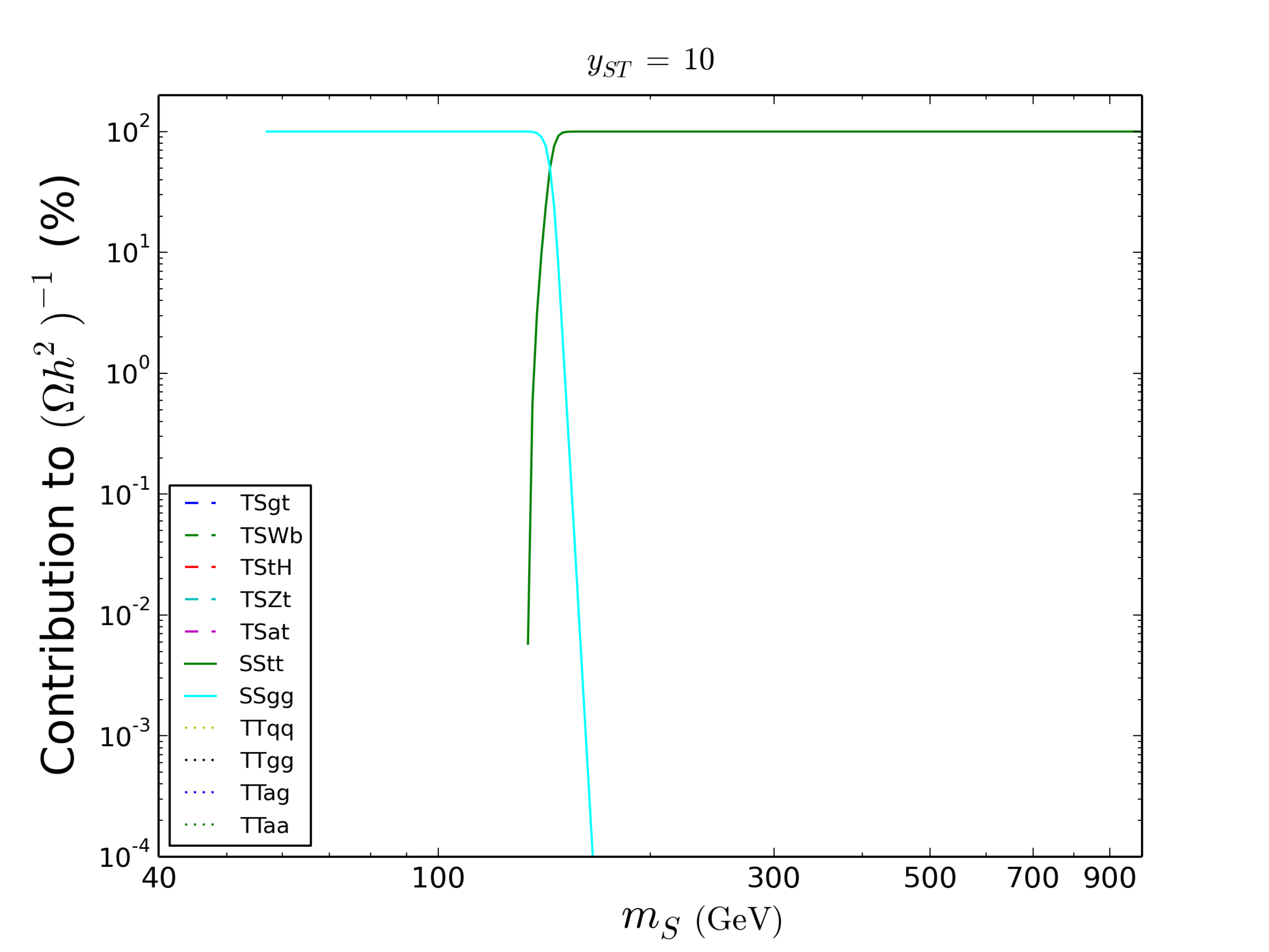}}
\caption{Contributions to the DM annihilation from different channels, for $y_{ST}=0.3, 0.5, 1.0, 10$.}
\label{fig:fc}
\end{center}
\end{figure}

\subsection{\label{subsection_topVL_H}Interplay between TopVL and Higgs portal}

Now we study the interplay between the topVL and Higgs portal. The interference happens between the $t/u$-channel processes $SS\to T^*\to t\bar{t}$ in the topVL portal and the $s$-channel process $SS\to h^*\to t\bar{t}$ in the Higgs portal. However, considering that $SS\to h^*\to t\bar{t}$ only occupies a small branch fraction in the Higgs portal annihilation (below $10\%$, see fig.2 in ref \cite{Cline2013}), we would expect generally constructive
contributions to the total annihilation cross section from other channels provided by the Higgs portal.
For each model point on it in fig.\ref{fig:mdm_vs_r_topVL} with different DM mass $m_S$, we set the Higgs
portal coupling to be $\lambda_{SH}=\lambda_0(m_S)\, r_\lambda$ where $\lambda_0(m_S)$ is the proper $\lambda_{SH}$ in
the Higgs portal for $m_S$ to obtain the observed relic density, and $r_\lambda$ is chosen to be $0.1, 0.2, 0.5, 1.0$
to control the Higgs portal strength. If the modified relic density is larger than those in fig.\ref{fig:mdm_vs_r_topVL}
(which is 0.12), then there must be destructive interference from $SS\to t\bar{t}$ between the topVL and Higgs portal
resulting in a  decreased total annihilation cross section. However, if the relic density becomes smaller we can not claim the
interference is constructive since the Higgs portal also provides other channels which will increase the annihilation
cross section. Note that here we consider $y_{ST}\lesssim1$, which means the $C_{SSgg}$ contribution is
negligible in most cases, especially for the DM mass ranges discussed here ($m_t < m_S < 1 {\rm TeV}$).

\begin{figure}[h]
\begin{center}
\includegraphics[width=12cm]{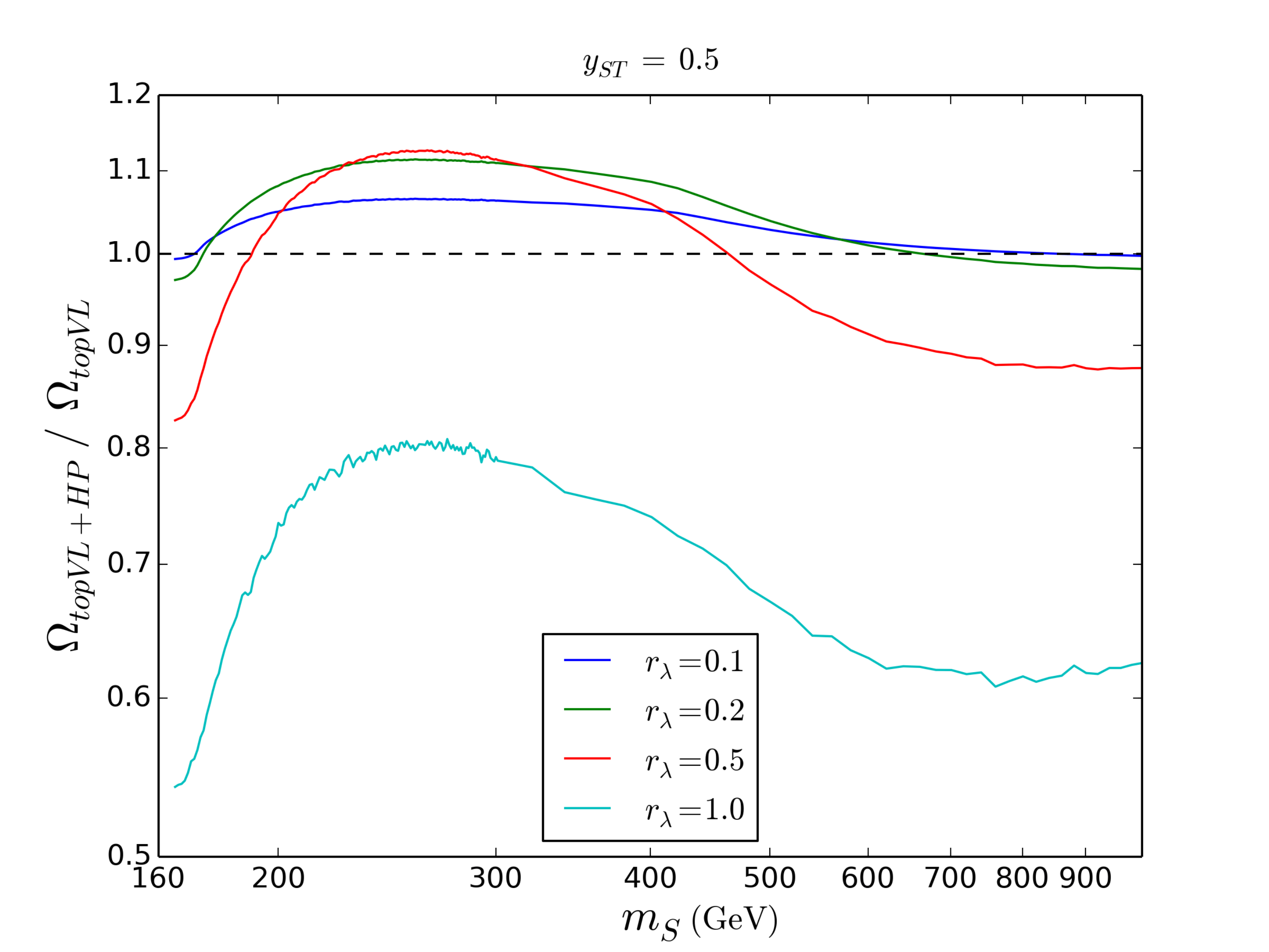}
\caption{Modified relic density for $y_{ST}=0.5$ curve in fig.\ref{fig:mdm_vs_r_topVL} by setting $\lambda_{SH}=\lambda_0(m_S)\, r_\lambda$ where $\lambda_0(m_S)$ is the proper $\lambda_{SH}$ in the Higgs portal for $m_S$ to obtain the observed relic density, while $r_\lambda$ is chosen to be $0.1, 0.2, 0.5, 1.0$. The vertical axis shows the ratio of the modified $\Omega_{\rm topVL+HP} h^2$ to $\Omega_{\rm topVL} h^2(=0.12)$ in fig.\ref{fig:mdm_vs_r_topVL}.}
\label{fig:topVL_H_inteference}
\end{center}
\end{figure}

\begin{figure}[h]
\begin{center}
\includegraphics[width=12cm]{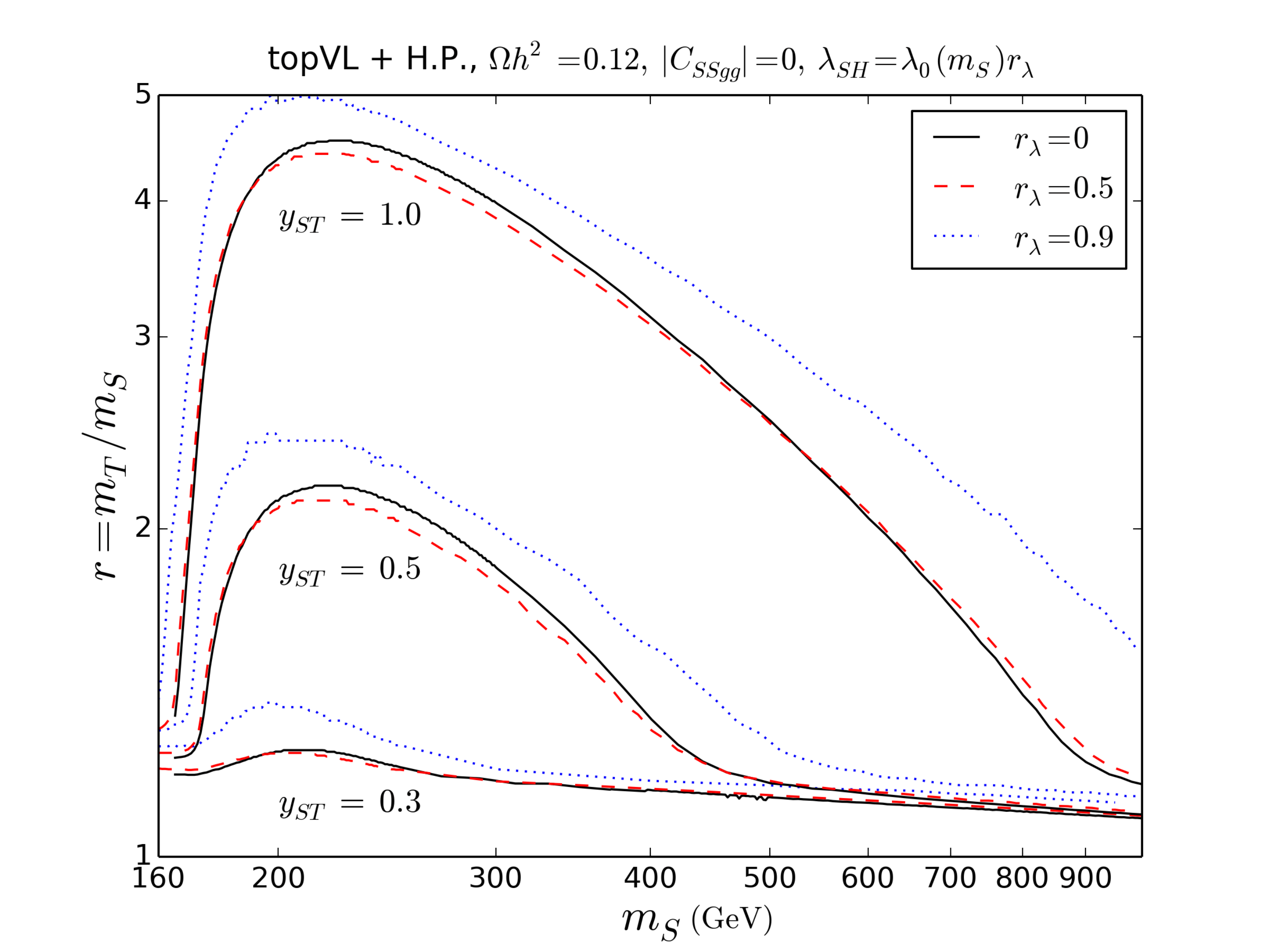}
\caption{Same as fig.\ref{fig:mdm_vs_r_topVL} but with Higgs portal contribution by setting $r_\lambda=0.5, 0.9$ to get the observed relic density.}
\label{fig:mdm_vs_r_topVL_H}
\end{center}
\end{figure}

In fig.\ref{fig:topVL_H_inteference} we show the modified relic density with $r_\lambda=0.1, 0.2, 0.5, 1.0$ for
$y_{ST}$=0.5. The vertical axis shows the ratio of $\Omega_{\rm topVL+HP} h^2$ to $\Omega_{\rm topVL} h^2$(=$0.12$) in
fig.\ref{fig:mdm_vs_r_topVL}. One can see that when $r_\lambda$ is small (e.g.  $0.1, 0.2, 0.5$) there are DM mass
wide ranges where $\Omega_{\rm topVL+HP}/\Omega_{\rm topVL} >1$ which means there exists destructive $SS\to t\bar{t}$
interference between the topVL and Higgs portal. However, for larger $r_\lambda$ the other annihilation channels in the
Higgs portal increases the total cross section significantly and results in an underproduced relic density.
Here we use the $s$-wave annihilation amplitude as an example to demonstrate the interference pattern, in which case one can set the relative velocity $v_{rel}$ between the two annihilating DM to be zero to simplify the calculation.
\begin{eqnarray}
 i \mathcal{M}_{t\bar{t}}
 &=& i\mathcal{M}_t + \, i\mathcal{M}_u + \, i\mathcal{M}_{H.P.} \\ \nonumber
 &=& \bar{u}_t (-i y_{ST} P_L)  \frac{i (\slashed{P}_1-\slashed{P}_t+m_T)}{(P_1 - P_t)^2-m_T^2} (-i y_{ST} P_R) v_{\bar{t}} \\ \nonumber
 &+& \bar{u}_t (-i y_{ST} P_L) \frac{i (\slashed{P}_2-\slashed{P}_t+m_T)}{(P_2 - P_t)^2-m_T^2} (-i y_{ST} P_R) v_{\bar{t}} \\ \nonumber
 &+& \bar{u}_t (-i \frac{m_t}{v}) \frac{i}{(P_1+P_2)^2-m_h^2} (-i\lambda v) v_{\bar{t}}
\label{eq:interference}
\end{eqnarray}
where $\mathcal{M}_{t\bar{t}}$ is the amplitude of annihilation into $t\bar{t}$ state which includes the $t/u$-channel from the topVL portal and the $s$-channel from the Higgs portal. $u_t, v_{\bar{t}}$ are the Dirac spinors of the top quark pair, and $P_L, P_R$ are the projection operators. $v\approx 246$ GeV is the vacuum expectation value in the SM. The momenta of the two scalar DM in the initial state are taken to be $P_1 = P_2 = (m_S, 0, 0, 0)$ since $s$-wave doesn't depend on the DM velocity. Under these simplifications and using the equation of motion of the top quark $\bar{u}_t(\slashed{P}_t-m_t)=0$, the above $\mathcal{M}_{t\bar{t}}$ can be simplified into
\begin{eqnarray}
 \mathcal{M}_{t\bar{t}}
 &\approx& - \bar{u}_t \Big[ 2y_{ST}^2 \frac{m_S (\gamma^0+r)-m_t}{m_t^2 - m_S^2 (1+r^2)}  +\lambda \frac{m_t}{4m_S^2 - m^2_h}  \Big] v_{\bar{t}} \\ \nonumber
 &\approx& \bar{u}_t \Big[2y_{ST}^2 \frac{m_S (\gamma^0+r)-m_t}{m_S^2 (1+r^2)} - \frac{\lambda}{4} \frac{m_t}{m_S^2}  \Big] v_{\bar{t}} 
\end{eqnarray}
with mass ratio $r=m_T/m_S$ defined previously. When $SS\to t\bar{t}$ is kinematically open with $m_S > m_t$, one can clearly see the opposite sign between these two portals which causes the destructive interference. Meanwhile, the mass ratio $r$ varying with $m_S$ shown in fig.\ref{fig:mdm_vs_r_topVL} also determines the interference strength and pattern in fig.\ref{fig:topVL_H_inteference} as $m_S$ increases. Nevertheless, since $SS\to t\bar{t}$ only occupies a small branch fraction $(<10\%)$ in the Higgs portal \cite{Cline2013}, other annihilation channels in the Higgs portal when $r_\lambda$ becomes larger will increase the final $\langle \sigma v \rangle$ and produce a reduced relic abundance.
These features can also be seen from fig.\ref{fig:mdm_vs_r_topVL_H} on the same plane as fig.\ref{fig:mdm_vs_r_topVL}, but
including contributions from both topVL and Higgs portal to get the correct relic density. For relatively large
$r_\lambda$ (e.g. 0.9), the significant contribution from the Higgs portal requires the mass ratio $r$ in topVL portal
to further deviate from 1 in order not to annihilate too fast. For smaller $r_\lambda$ (e.g. 0.5), however, the
parameter shift is relatively small and $r$ can be reduced due to destructive interference.

\section{\label{section_DD}Direct Detection}

Since the DM direct detection in Higgs portal models has been studied intensively in the literatures, here we turn off
the Higgs portal scattering by setting $\lambda_{SH}$=0 and focus on the $C_{SSgg}$ loop induced scattering. The real
scalar DM-nucleon elastic scattering cross section in terms of Wilson coefficient based on DM-parton effective operators
can be found in \cite{Hisano2015}. Here we capture some of the relevant points.

We start with the effective Lagrangian of the interactions between the real scalar DM $S$ and partons
\begin{equation}
 {\mathcal L}_{\rm eff} = \sum_{p=q,g} C^p_S \mathcal{O}^p_S,
\label{eq:Leff-S-qg}
\end{equation}
with 
\begin{eqnarray}
\mathcal{O}^q_S &\equiv& m_q S^2 \bar{q}q,\nonumber \\
\mathcal{O}^g_S &\equiv& \frac{\alpha_s}{\pi} S^2 G^{A\mu\nu}G^A_{\mu\nu}.
\label{eq:CqCg}
\end{eqnarray}
where $\alpha_s$ is the strong couplings constant and $G^A_{\mu\nu}$ is the field strength tensor of the gluon field.
The spin-independent (SI) coupling of the real scalar $S$ with a nucleon can be defined as
\begin{equation}
 {\mathcal L}_{\rm SI}^{(N)} =f_N S^2 \bar{N}N,
\label{eq:Leff-S-N}
\end{equation}
with
\begin{equation}
 f_N/m_N = \sum_{q=u,d,s} C_S^q f^{(N)}_{T_q} - \frac{8}{9}C^g_S f^{(N)}_{T_G},
\label{eq:fN}
\end{equation}
where $f^{(N)}_{T_q}$ is the quark mass fraction defined as $f^{(N)}_{T_q}\equiv \langle N| m_q \bar{q}q | N \rangle /
m_N$ \cite{PhysRevD.81.014503,PhysRevD.87.034509,Alarcon:2011zs,Alarcon:2012nr} and $f^{(N)}_{T_G}\equiv 1-\sum\limits_{q=u,d,s}f^{(N)}_{T_q}$ \cite{SHIFMAN1978443}. Finally the SI scattering cross section of the real
scalar with the target nucleus with mass $m_{\rm tar}$ can be expressed as
\begin{equation}
\sigma=\frac{1}{\pi}(\frac{m_{\rm tar}}{m_S+m_{\rm tar}})^2 | n_p f_p +n_n f_n |^2.
\label{eq:cs}
\end{equation}
Since there is no valence top quark in the nucleon, we only need to consider the gluon contribution here. Consequently, the loop coupling $C_{SSgg}$ plays a unique role in the direct detection when Higgs portal is turned off.

\begin{figure}[h]
\begin{center}
\includegraphics[width=12cm]{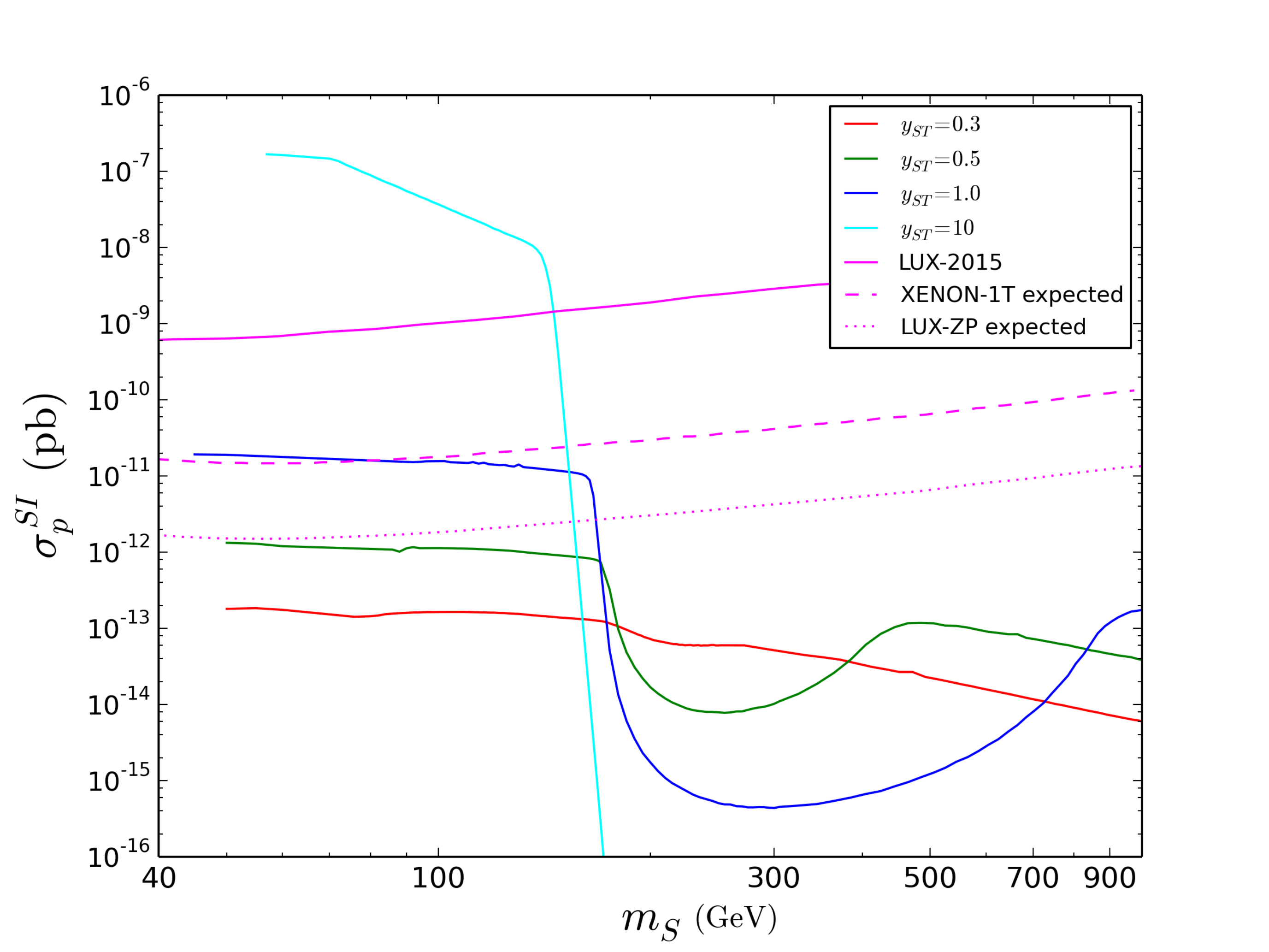}
\caption{The scalar DM-nucleon elastic scattering cross section in the direct detection coming from the $C_{SSgg}$ coupling, with $\lambda_{SH}$=0. The magenta solid, dash and dot lines are the current LUX bound \cite{LUXCollaboration2015} and anticipated sensitivity of XENON-1T and LUX-ZP \cite{DMtools}, respectively. The solid red, green, blue and cyan curves correspond to $y_{ST}$=0.3, 0.5, 1.0 and 10 which include the topVL and $C_{SSgg}$ contribution to obtain the observed relic density.}
\label{fig:DD}
\end{center}
\end{figure}

In fig.\ref{fig:DD} the magenta solid, dash and dot lines are the current LUX bound \cite{LUXCollaboration2015} and anticipated sensitivity of XENON-1T and LUX-ZP \cite{DMtools}, respectively. The solid red, green, blue and cyan curves correspond to $y_{ST}$=0.3, 0.5, 1.0 and 10 which include the topVL and $C_{SSgg}$ contribution to obtain the observed relic density.
One can see that the relaxed $r$ due to on-shell produced $SS\to t\bar{t}$ will suppress $C_{SSgg}$ and thus
$\sigma_p^{\rm SI}$, especially for large $y_{ST}=1.0, 10$ because they have larger $r$ (see
fig.\ref{fig:mdm_vs_r_topVL}). However, as discussed in section \ref{subsection_SSgg}, $SS\to gg$ with large $y_{ST}>1$
can increase rapidly and dominate over co-annihilation in some range of $m_S<m_t$ where $SS\to t\bar{t}$ is mostly
kinematically unavailable or inefficient. In this case, since the scattering process occurs via the crossed diagrams of the DM
annihilation, $C_{SSgg}$ (and thus $f_N$) is independent of $y_{ST}$ and fixed to the proper value depending on
$m_S$ to obtain the observed relic density. One can clearly see that $y_{ST}=10$ makes $SS\to gg$ dominate in a wide
range of $m_S<m_t$ (see fig.\ref{fig:fc}) where $\sigma_p^{\rm SI}$ only depend on the DM mass (see eq.(\ref{eq:cs})). We
found that the current LUX results can exclude this $SS\to gg$ dominating scenario for any sufficiently large $y_{ST}$,
although the perturbative $y_{ST}\lesssim1$ is beyond the current LUX sensitivity. However, the future XENON-1T
experiment may be capable of detecting $y_{ST}\gtrsim1$ for DM mass below around 100 GeV, while the LUX-ZP experiment
may further cover the smaller $y_{ST}\gtrsim0.5$.

\section{\label{section_ID}Indirect Detection}

Recently the sensitivity of DM indirect detection has been close to the canonical thermal annihilation cross section. In
today's Universe, DM $S$ in our model mainly annihilate into $t\bar{t}$ when $m_S>m_t$, while $SS\to gg$ is the dominant
annihilation channel for $m_S<m_t$. Here we consider the updated results of Fermi gamma-ray observations of continuous
spectrum from dwarf galaxies \cite{Ackermann:2015zua} as well as the line spectrum from the Galactic center region
\cite{Ackermann:2015lka}. We do not consider the constraints from charged cosmic particles such as positron and
anti-proton due to the relatively large uncertainties of their propagation models.

We first recall some main points of the analysis method based on the results of dwarf galaxy observations
\cite{PhysRevLett.107.241303}. The number of photon events observed can be divided into two independent factors: one
corresponding to the particle physics process and one describing the astrophysical information of the dwarf
galaxies. The expected number of signal events can be expressed as
\begin{equation}
\mu_\gamma(\Phi_{\rm PP}) \equiv (A_{\rm eff} T_{\rm obs}) \Phi_{\rm PP} J,
\label{eq:ID-Nevent}
\end{equation}
where $A_{\rm eff}$ is detector's effective area and $T_{\rm obs}$ is the exposure time. The $J$ factor contains the astrophysical information of the DM distribution and is defined by
\begin{equation}
J\equiv  \int_{\Delta\Omega(\psi)} \int_\ell [\rho(\ell,\psi)]^2 d \ell d\Omega(\psi),
\label{eq:ID-J}
\end{equation}
where the integration is performed along the line of sight in a direction $\psi$ and over a solid angle $\Delta\Omega$.

For self-conjugate DM particles $\chi$ the particle physics part is defined as
\begin{equation}
\Phi_{\rm PP}\equiv \frac{\langle \sigma_A v \rangle}{8\pi m_\chi^2} \int^{m_\chi}_{E_{\rm th}} \sum_f B_f \frac{dN_f}{dE} dE,
\label{eq:ID-PP}
\end{equation}
where $m_\chi$ is the DM mass and $\langle \sigma_A v \rangle$ is the total velocity-averaged cross section of DM annihilation
into SM particles in today's Universe. The $f$ denotes the annihilation channels and $B_f$ their branching fractions. For a given
channel, $dN_f/dE$ is its own final gamma-ray spectrum and the integration from threshold energy $E_{\rm th}$ to
$m_\chi$ gives the total number of photons emitted $N_{\gamma,f}=\int^{m_\chi}_{E_{\rm th}} \frac{dN_f}{dE} dE$.

Since the constraint on $\langle\sigma v\rangle_{t\bar{t}}$ is not given in Fermi dwarf galaxies results
\cite{Ackermann:2015zua}, we converted the $b\bar{b}$ bound to $t\bar{t}$ using $\langle\sigma
v\rangle_{t\bar{t}}=\langle\sigma v\rangle_{b\bar{b}}\, N_{\gamma,b\bar{b}}/N_{\gamma,t\bar{t}}$ as done in
\cite{Bringmann:2012vr,Giacchino:2015hvk}. The constraints on $\langle\sigma v\rangle_{gg}$ are obtained in \cite{Bringmann:2012vr,Giacchino:2015hvk} in a similar way. In our
model, both $t\bar{t}$ and $gg$ channels will contribute to the final gamma-ray spectrum, thus both the $\langle\sigma
v\rangle_{t\bar{t}}$ and $\langle\sigma v\rangle_{gg}$ bounds will put constraints on the cross section
$\langle\sigma v\rangle_{t\bar{t}}+\langle\sigma v\rangle_{gg}$. We also notice that the contribution from $SS\to
t\bar{t}\gamma$ is always negligibly small, which is different from the light quark scenario in \cite{Giacchino:2015hvk}.

As for the implementation of line spectrum observations, constraints can be obtained on $\langle\sigma
v\rangle_{\gamma\gamma}$ which is generated from the same diagram as the effective $SSgg$ coupling by replacing $g$ with
$\gamma$. The ratio of cross section $\langle\sigma v\rangle_{\gamma\gamma}/\langle\sigma v\rangle_{gg}$ is
given by \cite{Chu:2012qy}
\begin{equation}
\frac{\langle\sigma v\rangle_{\gamma\gamma}}{\langle\sigma v\rangle_{gg}} = \frac{9}{2} Q_t^4 (\frac{\alpha_{em}}{\alpha_s})^2 \approx 3.8\times10^{-3},
\label{eq:ID-ratio-aagg}
\end{equation}
where $Q_t$ is the top quark electric charge in term of $|e|$.

\begin{figure}[h]
\begin{center}
\subfigure{\includegraphics[width=7.5cm]{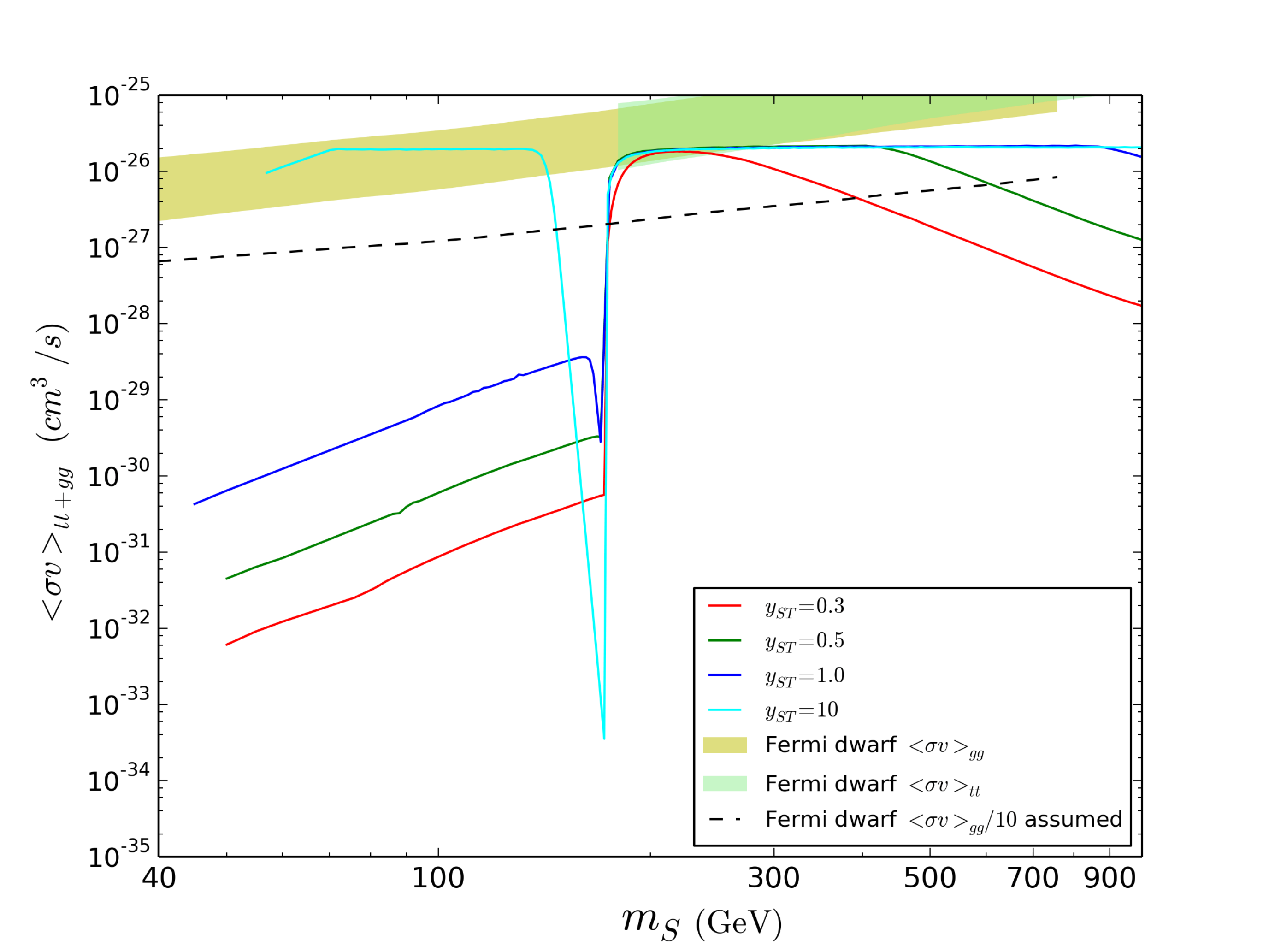}}
\subfigure{\includegraphics[width=7.5cm]{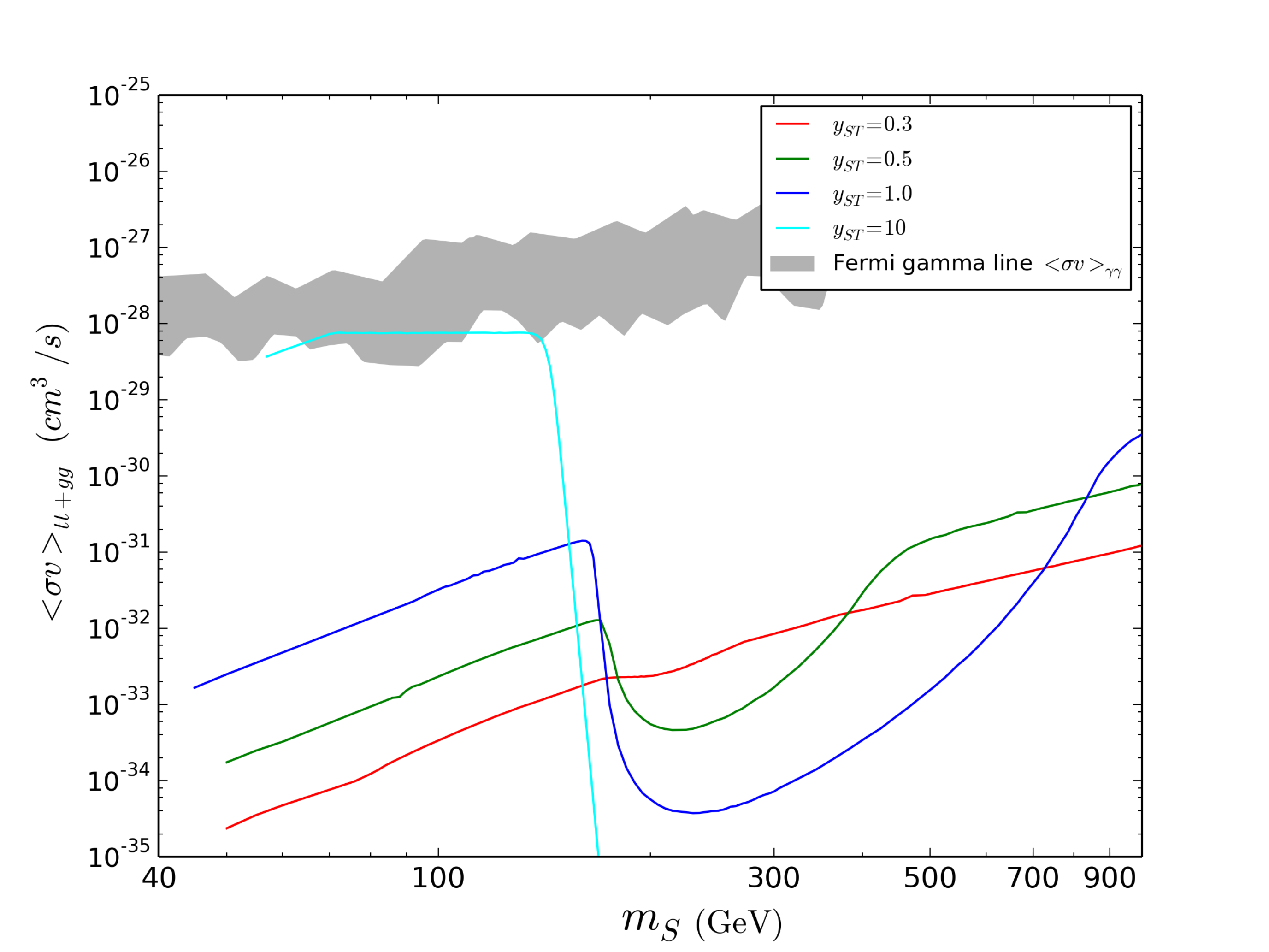}}
\caption{Indirect detection constraints from Fermi gamma-ray
  observations of continuous spectrum from dwarf galaxies \cite{Ackermann:2015zua} (left panel) and the line spectrum from the
  Galactic center region \cite{Ackermann:2015lka} (right panel). The bands reflect the uncertainties in the obtained bounds from the modeling of DM halo profile imposed in the Fermi reports \cite{Ackermann:2015zua,Ackermann:2015lka}. The samples plotted are the same as in fig.\ref{fig:fc} with $y_{ST}$=0.3, 0.5, 1.0, 10.0 which include the topVL and $C_{SSgg}$ contribution to obtain the observed relic density.}
\label{fig:ID}
\end{center}
\end{figure}

In fig.\ref{fig:ID} we show the indirect detection constraints on the samples of fig.\ref{fig:fc} which include the topVL
and $C_{SSgg}$ contribution to obtain the observed relic density. In both panels, the bands reflect the uncertainties in the obtained bounds from the modeling of DM halo profile imposed in the Fermi reports \cite{Ackermann:2015zua,Ackermann:2015lka}.
Similar to the LUX bound in direct detection, current
Fermi results from both dwarf and Galactic center observations can cover $SS\to gg$ dominating
scenario and exclude some DM mass range depending on the chosen DM profile. Moreover,  given the fact that the limits from Fermi-LAT based on 6 years data \cite{Ackermann:2015zua} (see fig.1 therein) have increased by an order of magnitude compared to 4 years data \cite{Ackermann:2015tah_1501.05464} and it is expected that Fermi-LAT will keep on accumulating data in the next two years claimed by the official website \cite{Fermi-LAT-Website}, we are motivated to consider the future sensitivity improvement by one order of magnitude. Based on this assumption, we can see a large part of the $SS\to t\bar{t}$ dominating region may also be excluded, pushing DM mass to be heavier than about 400, 600, 1000 GeV for $y_{ST}=0.3, 0.5, 1.0$,
respectively. For light DM mass $m_S<m_t$, however, perturbative $y_{ST}\lesssim 1$ can easily evade the constraints
from the current gamma-ray observation.

\section{\label{section_collider}Collider Search}

Since the top partner $T$ carries the color charge, $pp \to T\bar{T}$ can have sizable production cross section at the
LHC. The $T\bar{T}$  pair will decay through on-shell or off-shell top quarks plus DM particles which finally result in hadronic or leptonic final states with missing energy. In the collider study of our model, we considered the latest ATLAS 13.2 $\rm{fb^{-1}}$ data at 13 TeV of stop searches with $1\ell + jets + E^{\rm miss}_T$ signals \cite{ATLAS-CONF-2016-050} which shows an improvement (up to $m_{\tilde{t}_1}\sim 850$ GeV) in the exclusion capability compared to the 8 TeV 20.3 $\rm{fb^{-1}}$ data (up to $m_{\tilde{t}_1}\sim 710$ GeV), especially in the $m_{\tilde{t}_1} - m_{\tilde{\chi}^0_1} > m_t$ region with small $m_{\tilde{\chi}^0_1}$. Since the decay chain of our model is similar to the stop case and the production cross section of fermionic particle $T$ is generally larger than the scalar $\tilde{t}_1$, we would expect an even higher excluded $m_T$. 

We use FeynRules \cite{Alloul2014} to implement this top-philic model into MadGraph5 \cite{Alwall2014} to generate the parton level
events, followed by PYTHIA6 \cite{Sjostrand2006} to perform the parton shower. Then we use CheckMate
\cite{Kim2015,Drees2013}, which has encoded Delphes \cite{Ovyn2009,DeFavereau2014}, to simulate the collider response and obtain the cut efficiency $\epsilon$. Then the number of signal events are calculated as $N_{sig}=\mathcal{L}*\sigma*\epsilon$ where $\mathcal{L}=13.2 \,{\rm fb^{-1}}$ is the ATLAS integrated luminosity at 13 TeV in \cite{ATLAS-CONF-2016-050} and $\sigma$ is the production cross section of $pp\to T\bar{T}$ at 13 TeV. We use top++2.0 \cite{Czakon:1112-5675} to calculate $\sigma(pp\to T\bar{T})$ at next-to-next-to-leading order (NNLO) including also the next-to-next-to-leading logarithmic (NNLL) contributions. We vary the factorization and renormalization scale between $(0.5, 2)m_T$ to estimate the $1\sigma$ theoretical uncertainty $\Delta \sigma$. CheckMate will use this $\Delta \sigma$ and the number of generated simulation events $N_{MC}$ to calculate the total uncertainty of signal event number $\Delta N_{sig}$. Then CheckMATE defines the following quantity:
\begin{eqnarray}
r_{CM} \equiv \frac{N_{sig} - 1.96 \Delta N_{sig}}{N^{95}_{obs}}
\end{eqnarray}
where $N_{obs}^{95}$ is the model independent limits at $95\%$ Confidence Level (C. L.) on the number of new physics signal events given in the experimental reports. Then a model can be considered to be excluded at the $95\%$ C. L. if $r_{CM}>1$. This $r_{CM}$-limit is usually weaker than the usual method based on $S/\sqrt{S + B} < 1.96$ since $r_{CM}$-limit uses the total uncertainty on the $N_{sig}$ in a more conservative manner. More details can be found in \cite{Kim2015,Drees2013}.

There are seven Signal Regions (SRs) defined in \cite{ATLAS-CONF-2016-050}, of which the $\bm{\mathrm{SR1}}$ and $\bm{\mathrm{tN\_high}}$ are directly relevant to our case due to the similarity between our process $pp\to T\bar{T}, \, T\to tS$ and the stop case $pp\to \tilde{t}_1\tilde{t}_1^*, \tilde{t}_1\to t \tilde{\chi}^0_1$. While both assuming $100\%$ branching fraction (Br), $\bm{\mathrm{SR1}}$ focuses on small mass splitting between $\tilde{t}_1$ and $\tilde{\chi}^0_1$ in which case the decay products are fully resolved, while $\bm{\mathrm{tN\_high}}$ targets larger mass splitting leading to highly boosted top quarks and close-by jets. Since the first step of decay products are $t\bar{t}+E^{\rm miss}_T$, the dominant SM background processes include $t\bar{t}, Wt, t\bar{t}+Z(\to \nu\bar{\nu})$ and $W+jets$. 
And because all SRs defined in \cite{ATLAS-CONF-2016-050} are required to have exactly one signal lepton, for the $W$ bosons produced in the $t\bar{t},Wt$ events in the considered SRs, they can both decay leptonically with one of the two leptons being ¡®lost¡¯ (including not identified, not reconstructed, or removed in the overlap removal procedure), or one of them decays leptonically and the other decays through a hadronically decaying $\tau$ lepton. Other smaller SM backgrounds include di-bosons, $t\bar{t} +W, Z+jets$ and multijet events.

We first checked the reliability of our implementation of \cite{ATLAS-CONF-2016-050} into CheckMATE. We chose several supersymmetry (SUSY) samples on the exclusion bound of \cite{ATLAS-CONF-2016-050} and compare  $N_{sig} = \mathcal{L}*\sigma*\epsilon$ we calculated to the $N_{obs}^{95}$ given in \cite{ATLAS-CONF-2016-050}.  Here $\sigma(pp\to \tilde{t}_1\tilde{t}_1^*)$ is calculated by Prospino2 \cite{Beenakker:1996ed} and the cut efficiency $\epsilon$ is obtained from CheckMATE. To be consistent among different models, here we used only the $\sigma$ and $\epsilon$ to calculate a center value $N_{sig}$ for comparison with $N_{obs}^{95}$ while neglecting the $\Delta N_{sig}$ which depends on $N_{MC}$ and $\Delta \sigma$ since they may be quite different when switching from SUSY studies to other new models. Since \cite{ATLAS-CONF-2016-050} does not provide the detailed cut-flow information, we would consider our implementation to be reliable if our $N_{sig}$ is close to $N_{obs}^{95}$ in several different SRs simultaneously.

\begin{figure}[ht]
\begin{center}
\includegraphics[width=13cm]{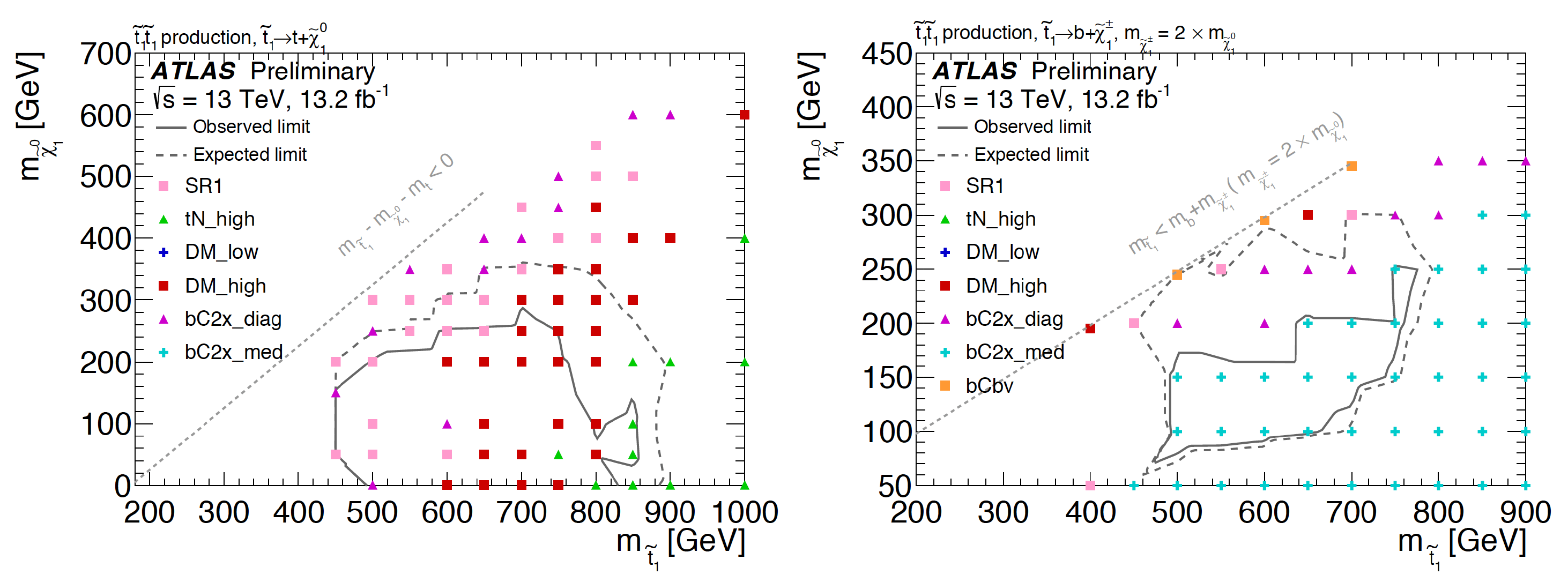}
\caption{Fig.15 of the latest ATLAS 13.2 $\rm{fb^{-1}}$ analysis with $1\ell + jets + E^{\rm miss}_T$ signals \cite{ATLAS-CONF-2016-050} we use for our validation.}
\label{fig15borrowed}
\end{center}
\end{figure}

Here we borrow fig.15 of \cite{ATLAS-CONF-2016-050} and show it in fig.\ref{fig15borrowed}. Although only the left panel of fig.15 in \cite{ATLAS-CONF-2016-050} has the decay chain $Br(\tilde{t}_1\to t \tilde{\chi}^0_1)=100\%$ mimicking our case $Br(T\to tS)=100\%$, we also considered the right panel of fig.15 for additional check of our implementation. Our validation results are shown in table.\ref{table-Validation} and the largest relative differences are about $20\%$ due to the small quantity $N^{95}_{obs}$. However, for $\bm{\mathrm{SR1}}$ and $\bm{\mathrm{tN\_high}}$ which directly apply to our decay chain $Br(T\to tS)=100\%$, the difference is $-13\%$ with very small $m_{\tilde{\chi}^0_1}$ and $1.5\%$ for moderate $m_{\tilde{\chi}^0_1}$. We did not consider fig.16 in \cite{ATLAS-CONF-2016-050} since the decay chains there are quite different from our model. We did not consider the fig.17 in \cite{ATLAS-CONF-2016-050} either, since they focus on the searches for new (pseudo-)scalar produced through fermion fusion $pp \to \phi(A) + t\bar{t} \to \chi\bar{\chi} + t\bar{t}$ where $\phi(A)$ is a new (pseudo-)scalar.

\begin{table}[ht]
\centering
\begin{tabular}{|c|c|c|c|c|c|}
\hline
SR        & panel of fig.\ref{fig15borrowed} &  $(m_{\tilde{t}_1}, m_{\tilde{\chi}^0_1})$ (GeV) & $N_{obs}^{95}$ & $N_{sig}$ & $(N_{sig}-N_{obs}^{95})/N_{obs}^{95}$ ($\%$)\\ \hline\hline
$\bm{\mathrm{SR1}}$        & left                               & (650,250)                                                                                 & 26             & 26.4      & 1.5                                                                                         \\ \hline
$\bm{\mathrm{tN\_high}}$   & left                               & (820,1)                                                                                   & 7.2            & 6.25      & -13                                                                                         \\ \hline
$\bm{\mathrm{bC2x\_diag}}$ & right                              & (650,350)                                                                                 & 12.4           & 12.4      & 0.0                                                                                         \\ \hline
$\bm{\mathrm{bC2x\_med}}$  & right                              & (650,200)                                                                                 & 9.9            & 10.8      & 9.1                                                                                         \\ \hline
$\bm{\mathrm{bCbv}}$       & right                              & (600,296)                                                                                 & 7.3            & 5.8       & -21                                                                                         \\ \hline
\end{tabular}
\caption{Validation of fig.15 of \cite{ATLAS-CONF-2016-050} (fig.\ref{fig15borrowed} in this paper) for five samples in five different signal regions, where $\bm{\mathrm{SR1}}$ and $\bm{\mathrm{tN\_high}}$ are the most important two which have the decay chain $Br(\tilde{t}_1\to t \tilde{\chi}^0_1)=100\%$ mimicking our case $Br(T\to tS)=100\%$.}
\label{table-Validation}
\end{table}

\begin{figure}[h]
\begin{center}
\includegraphics[width=12cm]{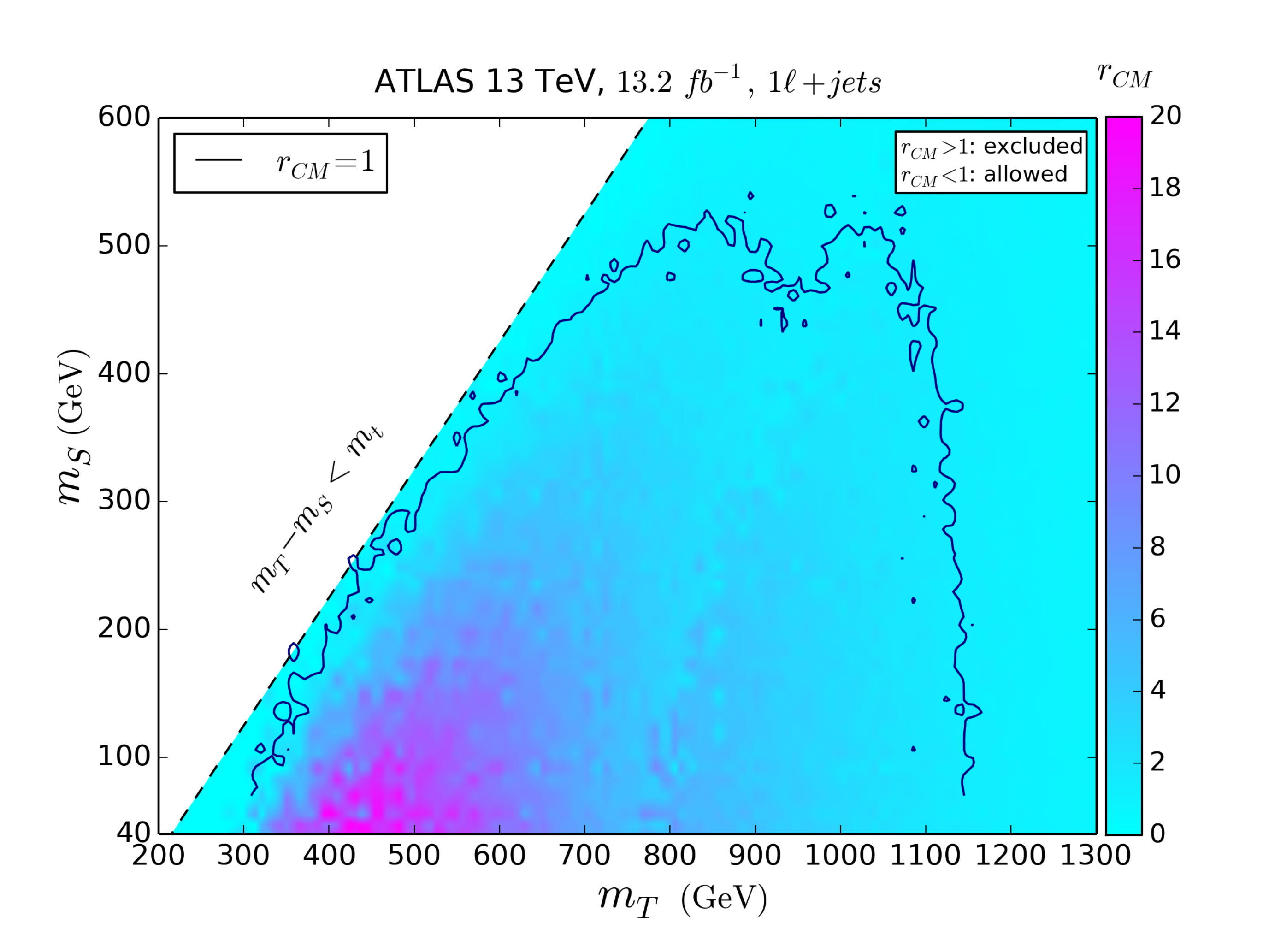}
\caption{$r_{CM}=(N_{sig} - 1.96 \Delta N_{sig})/N^{95}_{obs}$ calculated by CheckMate which can be used to claim a conservative exclusion at $95\%$ C.L. if $r_{CM}>1$.}
\label{fig:sstt_r}
\end{center}
\end{figure}

Now we turn to our top-philic model and in fig.\ref{fig:sstt_r} we show the $r_{CM}$-limit calculated by CheckMate where the black contour 
indicates $r_{CM}=1$. The region inside (outside) the $r_{CM}=1$ contour will be considered to be excluded (allowed) at $95\%$ C. L. by \cite{ATLAS-CONF-2016-050}.
The colored region satisfies $m_T - m_S > m_t$ which produces on-shell top quarks in the decay chain and is also the region studied in fig.15 of \cite{ATLAS-CONF-2016-050} (fig.\ref{fig15borrowed} in this paper).
We found that the latest ATLAS 13 TeV search can exclude a wide range of $m_T$ between 300
(650) and 1150 (1100) GeV for $m_S$ =40 (400) GeV and has the exclusion capability up to $m_S\sim 500$ GeV for this top-philic DM model.
This is an obviously wider region compared to the constraints on SUSY stop, where $m_{\tilde{t}_1}$ ($m_{\tilde{\chi}^0_1}$) up to 850 (250) GeV can be covered. 
We expect that the ongoing LHC Run-2 accumulating more data will extend this boundary.

\section{\label{section_combine}Combined Constraints on the Model}

\begin{figure}[ht]
\begin{center}
\subfigure{\includegraphics[width=10.5cm]{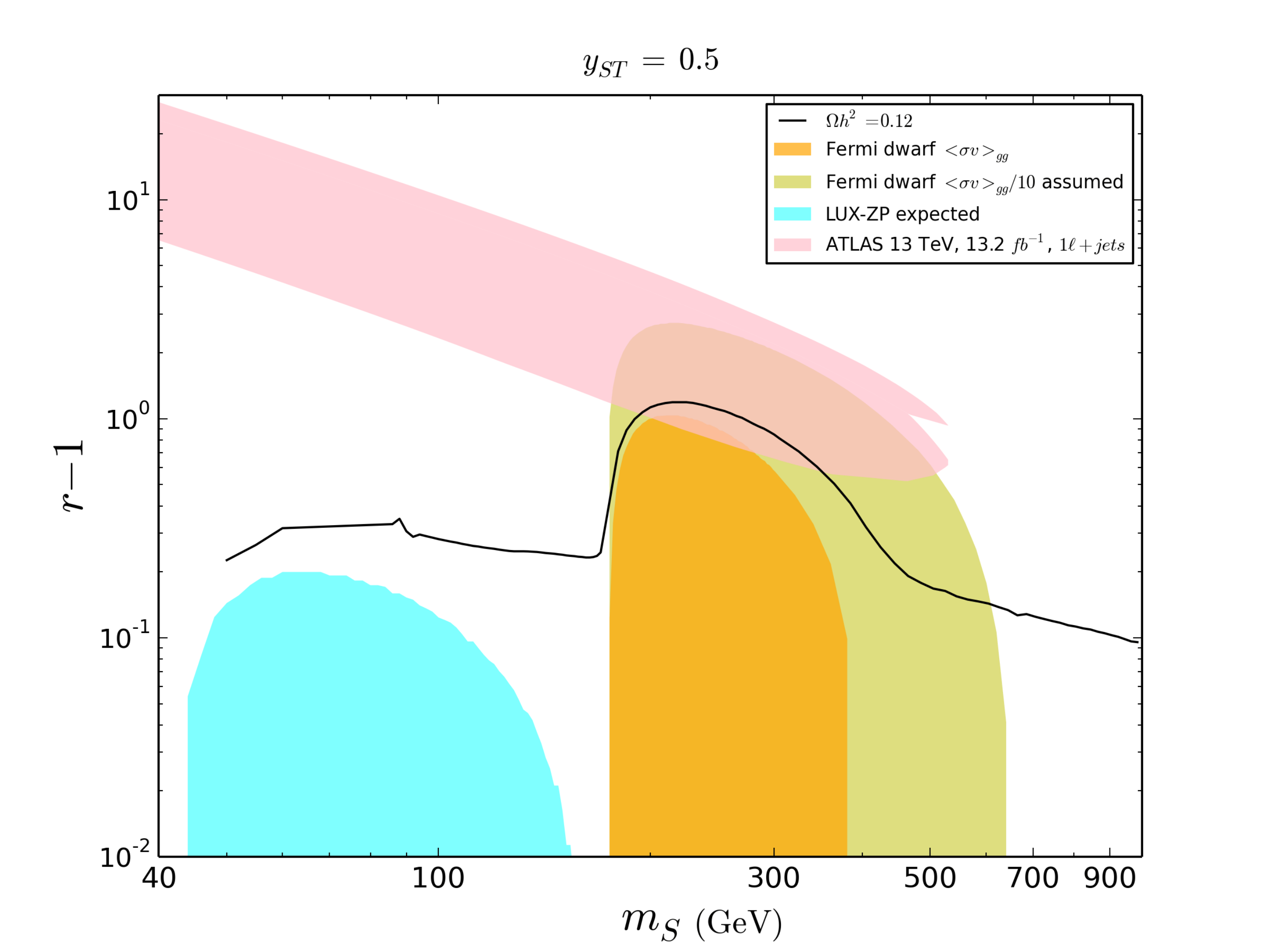}}
\subfigure{\includegraphics[width=10.5cm]{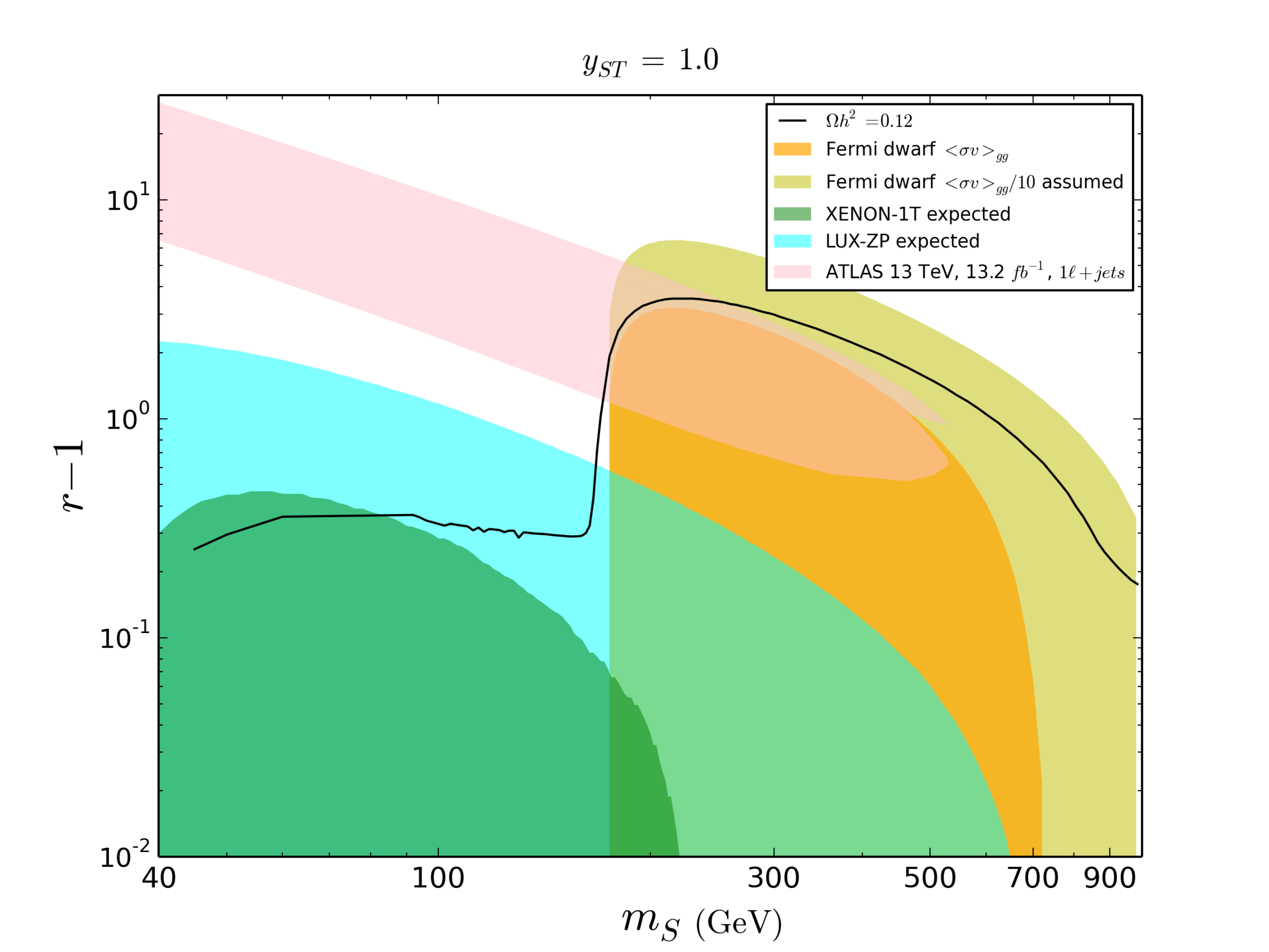}}
\caption{Combined results from thermal relic density, direct detection, indirect detection and collider search for $y_{ST}=0.5, 1.0$, where the black solid lines include the topVL and $C_{SSgg}$ contribution to obtain $\Omega_{\rm DM} h^2=0.12$. The light pick and orange regions correspond to the excluded parameter space by ATLAS 13 TeV 13.2 $\mathrm{fb^{-1}}$ data \cite{ATLAS-CONF-2016-050} and Fermi dwarf results $\langle\sigma v\rangle_{gg}$, while grey yellow and green (cyan) regions correspond to the assumed Fermi dwarf 10 times improvement $\langle\sigma v\rangle_{gg}/10$ and future XENON-1T (LUX-ZP) experiments.}
\label{fig:combine}
\end{center}
\end{figure}

Finally we combine the results from thermal relic density, direct detection, indirect detection and collider search in
fig.\ref{fig:combine}, where we choose $y_{ST}=0.5, 1.0$ given their perturbativity and possibility of detection
indicated from both fig.\ref{fig:DD} and fig.\ref{fig:ID}. The black solid lines include the topVL and $C_{SSgg}$
contribution to obtain $\Omega_{\rm DM} h^2=0.12$. The light pink, orange regions correspond to the excluded parameter
space by ATLAS 13 TeV 13.2 $\mathrm{fb^{-1}}$ data \cite{ATLAS-CONF-2016-050} and Fermi dwarf results $\langle\sigma v\rangle_{gg}$, while grey yellow
and green (cyan) regions correspond to the assumed Fermi dwarf 10 times improvement $\langle\sigma v\rangle_{gg}/10$ and future XENON-1T (LUX-ZP)
experiments. One can clearly see the complementarity between direct and indirect detection in the light and heavy DM
mass range, while the collider search result is independent of $y_{ST}$ since the top partner $T$ has only one decay
mode $T\to St$. We expect that a large portion of the parameter space will be covered by both the future direct and indirect experiments.

\section{\label{section_conclusion}Conclusion}

In this work we studied a scalar top-philic DM $S$
coupling, apart from the Higgs portal, exclusively to the right-handed top quark $t_R$ and a colored vector-like top
partner $T$ with Yukawa coupling $y_{ST}$ which we call the topVL portal. When the Higgs portal is closed and $y_{ST}$ is perturbative $ (\lesssim 1)$, $TS\to (W^+b, gt)$, $SS\to t\bar{t}$ and $T\bar{T}\to (q\bar{q},gg)$ provide the dominant contributions to obtain $\Omega_{\rm DM} h^2\simeq 0.12$  in light, medium and heavy DM mass range, respectively. However, large $y_{ST}\sim\mathcal{O}(10)$ can make $SS\to gg$ dominate via the loop-induced coupling $C_{SSgg}$ in the $m_S<m_t$ region.

Due to the absence of valence top quark in the nucleon, in this model it is the $C_{SSgg}$ coupling that generates DM-nucleon scattering which can be large when $SS\to gg$ dominates the DM annihilation. We found that the the current LUX results can exclude the $SS\to gg$ dominating scenario. The expected sensitivity of XENON-1T may further test $y_{ST}\gtrsim 1$, and $0.5\lesssim y_{ST}\lesssim 1$ may be covered in the future LUX-ZP experiment.

The indirect detection can play a complementary role in this model. The current results from Fermi gamma-ray
observations on both continuous spectrum from dwarf galaxies and line spectrum from Galactic center 
can also exclude the $SS\to gg$ dominating scenario
, and are just about to test the heavy DM mass region $m_S>m_t$. One order of magnitude of sensitivity improvement
can push DM mass to be heavier than about 400, 600, 1000 GeV for $y_{ST}=0.3, 0.5, 1.0$, respectively.

The colored top partner $T$ can be produced in pair at the hadron colliders such as
LHC. They will decay $100\%$ into $t\bar{t}+\slashed{E}_T$ signal when kinematically open and receive constraints from
the latest ATLAS 13 TeV 13.2 $\mathrm{fb^{-1}}$ data. We found that $m_T$ can be excluded between 300
(650) and 1150 (1100) GeV for $m_S$ =40 (400) GeV and the exclusion region can reach up to $m_S\sim 500$ GeV. We expect the
ongoing LHC Run-2 accumulating more data will extend this boundary.

\section*{Acknowledgement}

Peiwen Wu would like to thank Liangliang Shang, Yang Zhang and Yilei Tang for helpful discussions. We thank the Korea Institute for
Advanced Study for providing computing resources (KIAS Center for Advanced Computation Abacus System) for this work.
This work is supported in part by National Research Foundation of Korea (NRF) Research Grant NRF-2015R1A2A1A05001869 (SB,PK).

\bibliographystyle{JHEP}
\bibliography{SDM}

\end{document}